# Inducing ferromagnetism by structural engineering in a strongly spin-orbit coupled oxide

*Ji Soo Lim[1*], Carmine Autieri[2,3*], Merit Spring[1], Martin Kamp[1], Amar Fakhredine[4], Pavel Potapov[5], Daniel Wolf[5], Sergii Pylypenko[5], Axel Lubk[5,6], Johannes Schultz[5], Nicolas Perez[5], Börge Mehlhorn[5,6], Louis Veyrat[7], Mario Cuoco[3], Fadi Choueikan[8], Philippe Ohresser[8], Bernd Büchner[5,6], Giorgio Sangiovanni[9], Ralph Claessen[1] and Michael Sing[1*]*

1. Physikalisches Institut and Würzburg-Dresden Cluster of Excellence ct.qmat, Universität Würzburg, 97074 Würzburg, Germany
2. International Research Centre MagTop, Institute of Physics, Polish Academy of Sciences, PL-02668 Warsaw, Poland
3. CNR-SPIN, UOS Salerno, I-84084 Fisciano (Salerno), Italy
4. Institute of Physics, Polish Academy of Sciences, PL-02668 Warsaw, Poland
5. Leibniz Institute for Solid State and Materials Research Dresden and Würzburg-Dresden Cluster of Excellence ct.qmat, 01069 Dresden, Germany.
6. Institute of Solid State and Materials Physics and Würzburg-Dresden Cluster of Excellence ct.qmat, TU Dresden, 01062 Dresden, Germany.
7. CNRS, Laboratoire National des Champs Magnétiques Intenses, Université Grenoble-Alpes, Université Toulouse 3, INSA-Toulouse, EMFL, 31400 Toulouse, France.
8. Synchrotron SOLEIL, L'Orme des Merisiers, 91190 Saint-Aubin, France
9. Institut für Theoretische Physik und Astrophysik and Würzburg-Dresden Cluster of Excellence ct.qmat, Universität Würzburg, 97074 Würzburg, Germany.

E-mail: ji.lim@kit.edu, autieri@magtop.ifpan.edu.pl, michael.sing@uni-wuerzburg.de

## Abstract

Magnetic materials with strong spin-orbit coupling (SOC) are essential for the advancement of spin-orbitronic devices, as they enable efficient spin-charge conversion, complex magnetic structures, spin-valley physics, topological phases and other exotic phenomena. 5d transition-metal oxides such as $SrIrO_3$ feature large SOC, but usually show paramagnetic behavior due to broad bands and a low density of states at the Fermi level, accompanied by a relatively low Coulomb repulsion. Here, we unveil ferromagnetism in 5d $SrIrO_3$ thin films grown on $SrTiO_3$ (111). Through substrate-induced structural engineering, a zigzag stacking of three-unit-cell thick layers along the [111] direction is achieved, stabilizing a ferromagnetic state at the interfaces. Magnetotransport measurements reveal an anomalous Hall effect below ~30 K and hysteresis in the Hall conductivity below 7 K, indicating ferromagnetic ordering. X-ray magnetic circular dichroism further supports these results.

Theoretical analysis suggests that the structural engineering of the $IrO_6$ octahedral network enhances the density of states at the Fermi level and thus stabilizes Stoner ferromagnetism. This work highlights the potential of structurally engineered 5d oxides for spin-orbitronic devices, where efficient control of SOC-induced magnetic phases by electric currents can lead to lower energy consumption and improved performance in next-generation device technologies.

## 1. Introduction

Magnetic materials play a crucial role in current information-processing devices such as hard disks and magnetic random-access memories (MRAMs), and they are actively being developed for future spin-orbitronic devices [1, 2, 3, 4, 5, 6]. In particular, there has been a growing interest in controlling the direction of magnetization using electric currents, as this offers significant potential for novel devices with low energy consumption [7, 8, 9]. A key factor for the realization of efficient spin-charge conversion is the presence of strong spin-orbit coupling (SOC), which increases the spin Hall angle and generates larger spin currents [10]. Ferromagnetic materials with strong SOC allow a perpendicular magnetic moment to be efficiently switched by an in-plane current flow without an external magnetic field [11, 12, 13].

To achieve large SOC, materials containing 4d or 5d heavy elements are usually required. However, these elements have extended orbitals, which lead to a delocalization of electrons that weakens electron-electron correlations and often results in paramagnetic behavior. Aside from conventional exchange interactions, ferromagnetism in such systems can also arise from satisfying the Stoner instability criterion, as exemplified by $SrRuO_3$ – a well-known 4d itinerant ferromagnet [14, 15]. Recently, it has been shown that applying an electric current can change its out-of-plane magnetization direction in thin films with (001)-orientation [16]. These findings highlight the importance of designing ferromagnetic materials with strong SOC for spin-orbitronic device applications.

Here, we report the emergence of ferromagnetism in thin films of 5d $SrIrO_3$ (SIO) with a $d^5$ electron configuration when grown with (111)-orientation on $SrTiO_3$. In contrast, epitaxially stabilized (001)-oriented SIO films, which are known to adopt a perovskite structure, exhibit paramagnetic, metallic behavior across various substrates. For instance, 35 nm-thick films grown on $GdScO_3$ (110) – which is nearly lattice-matched to the pseudocubic lattice derived from orthorhombic bulk SIO – were found to remain paramagnetic down to 300 mK [17]. On the other hand, films on $SrTiO_3$ (001) undergo a metal-insulator transition (MIT) at a critical thickness of four unit cells (uc), i.e., towards the 2D limit when correlations are enhanced. This may be accompanied by in-plane magnetism arising from canted antiferromagnetic (AFM) ordering, as suggested by density functional theory (DFT) calculations [18, 19]. However, their highly insulating nature would, in any case, limit their suitability for current-driven devices. Signatures of ferromagnetic phases were observed in artificial $[(SrIrO_3)_m,SrTiO_3]$ superlattices, but again only in conjunction with insulating behavior, below $m$=3 [20], while itinerant ferromagnetism, induced by proximity, was reported in superlattices with the ferromagnetic insulator $LaCoO_3$ [21]. Generally, in $d^5$ iridates, the large SOC leads to the entanglement of the spin and orbital degrees of freedom, associated with the formation of $J_{eff}$=1/2 states. This results in a narrow, half-filled band, susceptible to Mott localization and favors superexchange-driven AFM interactions, specifically for corner-sharing octahedra with 180° Ir-O-Ir bond angles. This typically stabilizes a robust antiferromagnetic insulating phase that is difficult to convert into a ferromagnetic state [22, 23].

Given that bulk SIO, which adopts a monoclinic distortion of the hexagonal $BaTiO_3$ structure, is a non-Fermi-liquid metal close to a ferromagnetic instability [24], SIO (111) thin films offer a promising route to stabilize a hexagonal lattice motif and potentially realize intrinsic itinerant ferromagnetism. Also, due to the geometrically frustrated spin configuration in combination with strain, unexpected magnetic states with potential topological characteristics could emerge [25]. Both perspectives highlight SIO (111) films as an intriguing materials platform well worth closer investigation. However, achieving perovskite SIO (111) films is known to be challenging, primarily due to the fact that the bulk monoclinic phase is thermodynamically more stable [26]. Recent studies have demonstrated that *monoclinic* SIO films can be epitaxially grown on STO (111) substrates [27, 28], featuring a thickness-driven transition into a magnetic insulating state [28]. Furthermore, superlattice engineering has been employed to realize a well-defined perovskite structure with (111) orientation. A SIO(6 uc)/STO(4 uc) superlattice exhibits an anomalous Hall effect under both in-plane and out-of-plane magnetic fields, with the magnetic easy axis lying in-plane [29]. While these studies provide intriguing insights into the magnetic state, detailed investigations of the crystal structure and the spin configuration of SIO (111) films are still lacking, and the link between structure and magnetism remains unclear. In this work, we combine structural and magnetotransport measurements with theoretical analysis to shed light on the magnetic properties and their microscopic origin in SIO (111) films.

## 2. Results and discussion

We first explore the structural properties of a 9 uc thick SIO (111) film grown on STO (111). The thickness of the SIO (111) film, fabricated by pulsed laser deposition, was determined from the reflection high-energy electron diffraction (RHEED) oscillations (s. Fig. S1a in Supporting Information). After finishing the film growth, we carried out *in-situ* low-energy electron diffraction (LEED), exhibiting a six-fold structural symmetry (see Supporting Information Fig. S1c). X-ray diffraction shows a single-phase SIO (111) film with a *c*-axis lattice parameter of about 2.35 Å (Fig. 1a). The width of the rocking curve of the SIO (111) peak is about 0.15° (see inset of Fig. 1a), indicating that the film is of high crystalline quality.

An additional peak between the SIO (111) and (222) peaks is due to a superstructure corresponding to a period of 3 uc, which is confirmed by the high-angle annular dark field (HAADF) scanning transmission electron microscope (STEM) image in Fig. 1b. The superstructure is formed by twinning, which leads to a zigzag geometric arrangement of Ir and Sr atoms along the c-axis (see Fig. 1b). The in-plane lattice parameter obtained from TEM images along different zone axes, namely $[1\bar{1}0]$ and $[11\bar{2}]$, shows that the film is well strained to the STO substrate (see Supporting Information, Fig. S2). To confirm the six-fold structural symmetry from the LEED measurement, we compare STEM images from three other lamellae, cut at different angles as sketched in Fig. 1c, top. From these data, we determined the Ir positions with pm-scale precision, which are marked by green, red, and blue dots, corresponding to the lamellae with angles of 0º, -60º, and 60º with respect to the $[11\bar{2}]$ axis, respectively (Supporting Information Fig. S2d). The Ir atomic positions from the different lamellae coincide well with each other, indicating a 6-fold symmetry as seen by LEED (Fig. 1c, bottom). This implies that the SIO (111) film shows a hexagonal superstructure. By combining all TEM images, we obtain the three-dimensional crystal structure depicted in Fig. 1d. The interface between twinned stacks exhibits a face-shared octahedral structure, whereas the interior of the stacks consists of corner-shared octahedra as in a perovskite structure. The average distance between Ir-Ir ions in the face-shared octahedral arrangement is determined to be 2.7 Å, whereas in the corner-shared octahedral arrangement, it measures 3.9 Å (Fig. 1e).

To explore the electric transport properties, we first investigate the sheet resistance of SIO (111) films of different thicknesses using van-der-Pauw geometry. The SIO (111) films exhibit a thickness-dependent MIT at a critical thickness of 5 uc (Fig. 2a). Below this critical thickness, the SIO (111) films only feature corner-shared octahedra and display insulating behavior. On the other hand, the SIO (111) films above the critical thickness begin to show twinning with a zigzag atomic arrangement and the formation of corresponding interfaces with face-shared octahedra. The interactions between these face-shared and the corner-shared octahedra are expected to alter the film's electronic band structure, likely contributing to its metallic behavior. In the following, we choose a 9 uc thick SIO (111) film comprising two interfaces with face-shared octahedra to exemplarily study its magnetoelectric transport properties $R(B)/R(0)$ (s. Fig. 2b), where $R(B)$ and $R(0)$ are the sheet resistances at a finite $B$-field and without $B$-field, respectively. The magnetoresistance $(R_{9T} - R_{0T})/R_{0T}$, plotted in Fig. 2c, shows a sudden change from an increasing to a decreasing behavior approximately at 20 K and a sign change from negative to positive at about 7 K. Below 7 K, in the low $B$-field regime (below ~ 1 T), in addition, a hysteresis loop appears, a typical hallmark of magnetic materials. If ferromagnetic domains in the SIO (111) films align with the field, the reduction of magnetic fluctuations/magnons when approaching saturation can lead to a negative magnetoresistance, as for instance in magnetic 3D topological insulators [30, 31, 32]. As the field increases and the magnetization saturates, a further increase in the field's strength leads to orbital effects caused by the Lorentz force, which bends the paths of charge carriers and increases scattering, thereby increasing the resistance [30]. This results in a crossover to positive MR at high magnetic fields, which is observable even at 2 K in a sufficiently strong field of 14 T (Supporting Information Fig. S5a). Noticeably, a comparison between the magnetoresistance of SIO (001) and (111) films (see Fig. 2c) reveals that, unlike the (111) films, SIO (001) films consistently exhibit positive MR at fields of ±9 T across the entire temperature range (see Fig. S3, Supporting Information) [33, 34, 35]. To further explore the electronic carrier properties in SIO (001) and (111) films, we conducted anomalous Hall effect (AHE) measurements. Unlike the well-studied SIO (001) film, which exhibits linear Hall resistance and electron-dominated transport down to low temperatures, the SIO (111) film shows hole-dominated transport across the entire temperature range (Supporting Information Fig. S4).

To further analyze the AHE, we isolate its contribution by subtracting a linear component, i.e., the ordinary Hall effect, in the region where AHE is observed. We plot the anomalous Hall conductivity in Fig. 2e using the equation $\sigma_{AHE} = \rho_{AHE}/({\rho_{AHE}}^2 + {\rho_{xx}}^2)$, where $\rho_{AHE}$ is the transverse resistivity corresponding to the AHE and $\rho_{xx}$ represents the longitudinal resistivity. The resulting conductivity reaches values up to about 2 $\Omega^{-1}cm^{-1}$, among the highest values reported for oxides and comparable to those in strained pyrochlore iridates such as $Nd_2Ir_2O_7$ [36, 37]. The AHE begins to show hysteresis around 7 K, becoming more pronounced as the temperature decreases. The coercive field of the hysteresis loops scales linearly with the logarithmic temperature (see Fig. 2f). This behavior is attributed to ferromagnetic domain motion, where the minimum coercive field required to overcome domain wall pinning is proportional to $\exp\left[-\left(\frac{k_B T}{U_i}\right)\right]$, with $k_B$ and $U_i$ being the Boltzmann constant and the characteristic energy, respectively [38].

The AHE arises from both intrinsic and extrinsic mechanisms: the intrinsic contribution stems from a finite Berry curvature, while the extrinsic part originates from the magnetic scattering of electrons in ferromagnetic materials [39]. In our data, the Hall conductivity exhibits saturation accompanied by hysteresis, consistent with ferromagnetic behavior. Notably, even prior to the onset of hysteresis, the Hall conductivity reaches values close to saturation between 10 and 25 K, suggesting that intrinsic effects contribute to the AHE alongside ferromagnetism.

At high magnetic fields, angle-dependent Hall measurements reveal unusual behavior in the in-plane configuration (Supporting Information Figs. S5b and c). While the AHE signal is typically expected to gradually vanish as the magnetic field tilts towards the sample surface, we observe a butterfly-shaped AHE under tilted magnetic fields. This likely results from the saturation field not being reached at the field orientations other than perpendicular to the surface, as it exceeds the maximum experimentally available field of 14 T. Nevertheless, the ferromagnetic-like hysteresis in the AHE under out-of-plane magnetic field indicates ferromagnetic ordering with a preferential out-of-plane magnetic anisotropy.

To independently verify ferromagnetism in SIO (111) films, we performed X-ray magnetic circular dichroism (XMCD) measurements at the Ir $M_5$ edge ($3d_{5/2} \rightarrow 5f$ or $6p$) in the energy range from 2040 to 2070 eV in normal incidence geometry at the DEIMOS beamline at the SOLEIL synchrotron facility (France) [40]. The 5f orbital has a larger radial extension than the 4f orbital, which results in stronger spatial overlap with the 5d orbital. This suggests that the 5f–5d interaction can be significantly stronger than the 4f–5d interaction. Although no calculations have been reported for Ir, studies have been conducted on typical 5f systems, i.e., actinide compounds. For example, in $ThO_2$ ($Th^{4+}$, $5f^0$ configuration), 5f and 5d orbitals share the same principal quantum number and have similar radial distributions, leading to substantial orbital overlap [41]. *Ab-initio* Hartree-Fock calculations yield large positive values for the Slater integrals of actinide ions, indicating a strong and positive exchange interaction [42, 43]. As a result, when an electron is excited into an empty 5f orbital, when its spin is parallel to the spin-polarized 5d valence electrons, the total energy of the system is lowered. Conversely, antiparallel spin alignment is resulting in a higher energy state.

Based on this, XMCD at the Ir $M_5$ edge provides evidence of magnetism. In Fig. 3a, we plot the absorption spectra for left- and right-circularly polarized light recorded at 5 K and 6 T, which only show subtle differences, indicating tiny magnetic moments, oriented perpendicular to the SIO (111) film surface. As shown in Fig. 3b, the XMCD signals for + 6 T and - 6 T are inverted relative to each other, which indicates that the magnetic moments align with the external field and change direction when the magnetic field is reversed. The normalized XMCD areas can be regarded as a measure of the magnetization. When plotted as a function of magnetic field over a cycle from -6 T to +6 T and back (see Fig. 3c and Supporting Information, Fig. S6, for the XMCD signals), a hysteresis-like loop as a characteristic of ferromagnetism is obtained. The extracted coercive field is about 0.5 T, which agrees well with the value that can be obtained from Fig. 2d at the corresponding temperature.

We further examine the temperature dependence of the normalized XMCD integral, i.e. the magnetization, after the SIO (111) film was magnetized at -6T, in zero field, i.e., in remanence, starting from 5 K up to 200 K (see Fig. 3d). One can clearly see an increase below 50 K (see also Supporting Information, Fig. S7), supporting that the appearance of MR and AHE approximately below 30 K, as discussed above, are connected to the emergence of a magnetic state. Above 50 K, the XMCD signals and thus the normalized integrals become too weak to be reliably evaluated, with values close to zero. This suggests that the out-of plane magnetization vanishes above this temperature.

To understand the nature and the origin of the ferromagnetic state of the SIO (111) films, we investigated the electronic band structure by DFT calculations including on-site Coulomb repulsion (DFT + U) and X-ray photoelectron spectroscopy (XPS). The crystal structure, which was used as input for the *ab initio* calculations, was constructed from the analysis of TEM images (Fig. 1), with considering spin orientations along the x-, y-, and z-axes. Also, both magnetic and nonmagnetic states were examined to determine the spin configuration with the lowest energy. Our DFT results show that the spins in the ground state are aligned along the z-axis, i.e., perpendicular to the surface, consistent with our experimental findings (Fig. 4a). The small energy differences between competing magnetic states imply a small transition temperature $T_C$.

To investigate whether the different structural connectivity of the octahedra in the 3uc thick layers and at their interfaces plays a role in the emergence of magnetism, we performed further analysis. Indeed, the DFT calculations show a higher density of states (DOS) at the Fermi energy $E_F$ associated with the Ir 5d-orbitals for model structures featuring both face- and corner-shared octahedra, as in our films. In contrast, in the case of orthorhombic $SrIrO_3$, which contains solely corner-shared octahedra, a dip at $E_F$ is found, indicative of a band gap opening caused by strong SOC (Fig. 4d). As a result, the latter situation does not fulfill the Stoner criterion and a paramagnetic metal is obtained, whereas in the former case, the larger DOS at $E_F$ does satisfy the Stoner criterion. Moreover, the existence of face-shared octahedra every 3 uc reduces the effective SOC, thereby preventing the system from turning into a Mott insulator [44]. Together, these effects lead to an itinerant ferromagnetic state, with the DFT calculations yielding a net magnetic moment of 0.03 $\mu_B$/Ir (Supporting Information, Fig. S8).

Further analysis of the partial density of states associated with the Ir 5d-orbitals reveals that both corner- and face-shared octahedra contribute to the large DOS at $E_F$, with the dominant contribution arising from the former (Fig. 4e). This suggests that primarily the d orbitals of the corner-shared octahedra drive the Stoner instability. On the other hand, the face-shared octahedra play a more significant role in reducing the effective SOC and enabling the structural rearrangement. Additional theoretical results – presented in the Supporting information, Figs. S9 and S10 – highlight the presence of semi-Dirac points without magnetization arising from non-symmorphic symmetries, the emergence of Weyl points with finite magnetization that enhance the AHE, and the Fermi surfaces.

To examine the role of face-shared octahedra with respect to the electronic band structure in light of our theoretical findings, we recorded XPS valence band spectra of SIO (001) and (111) films. The SIO (111) film comprises both face- and corner-shared octahedra, while the SIO (001) film consists solely of corner-shared octahedra. Indeed, the Ir 5d derived DOS near $E_F$ is found to be significantly higher for the SIO (111) film (Fig. 4b) in good agreement with the theoretical calculations in Fig. 4d. In addition, the shorter Ir-Ir distance within the interfaces of face-shared octahedra leads to a stronger orbital overlap than for the corner-shared octahedra and thus a slightly larger bandwidth (Fig. 4b). Valence band spectra for SIO (111) films with varying thickness (Fig. 4c) confirm this picture. If the thickness is too small for the film to form an interface with face-shared octahedra, i.e., at a thickness of 3 uc, the DOS is significantly reduced compared to the films with a thickness of 6 and 9uc. In addition, a MIT with a gap opening occurs, probably due to the electron confinement along the z direction and the associated reduction in the kinetic energy with respect to potential energy due to short-range Coulomb repulsion.

## 3. Conclusion

Our work presents a material system in which an out-of-plane magnetic state is realized in a strongly spin–orbit coupled oxide via structural engineering. In contrast to the commonly observed in-plane magnetism in SIO (111) films, we show that a distinct magnetic anisotropy arises when the bulk monoclinic SIO phase is strained to an STO (111) substrate. This strain stabilizes a hexagonal, twinned superstructure featuring coexisting face- and corner-shared $IrO_6$ octahedra, providing a new platform for engineering spin orientation in 5d transition-metal oxides.

These films exhibit an AHE with a pronounced hysteresis at low temperatures. The emergence of ferromagnetism is confirmed spectroscopically by XMCD. DFT + U calculations show that the underlying main mechanism is a Stoner instability driven by a large DOS at $E_F$, originating from the combined existence of corner- and face-shared octahedra. We argue that the shortened Ir-Ir distance in the interfaces with the face-shared octahedra weakens the effective SOC and enhances the DOS to promote Stoner ferromagnetism.

The strong SOC in this ferromagnetic compound generates a Berry curvature as evidenced by the pronounced AHE observed between 10 and 20 K without hysteresis. Band structure calculations further indicate the existence of Weyl points close to the Fermi energy, suggesting that the material may host a ferromagnetic Weyl semimetal phase.

In spin-orbitronic devices, the SOC is a crucial ingredient both for realizing the spin Hall effect and for manipulating the spin orientation by heterointerfacing with ferromagnets. In this context, our results on SIO(111) films are particularly compelling as they integrate strong SOC and ferromagnetism within the same material system. When incorporated into all-oxide architectures, SIO(111) layers have the potential to serve as an essential functional building block for spin-orbitronic devices, marking an important step forward in both fundamental research and the development of next-generation oxide electronics.

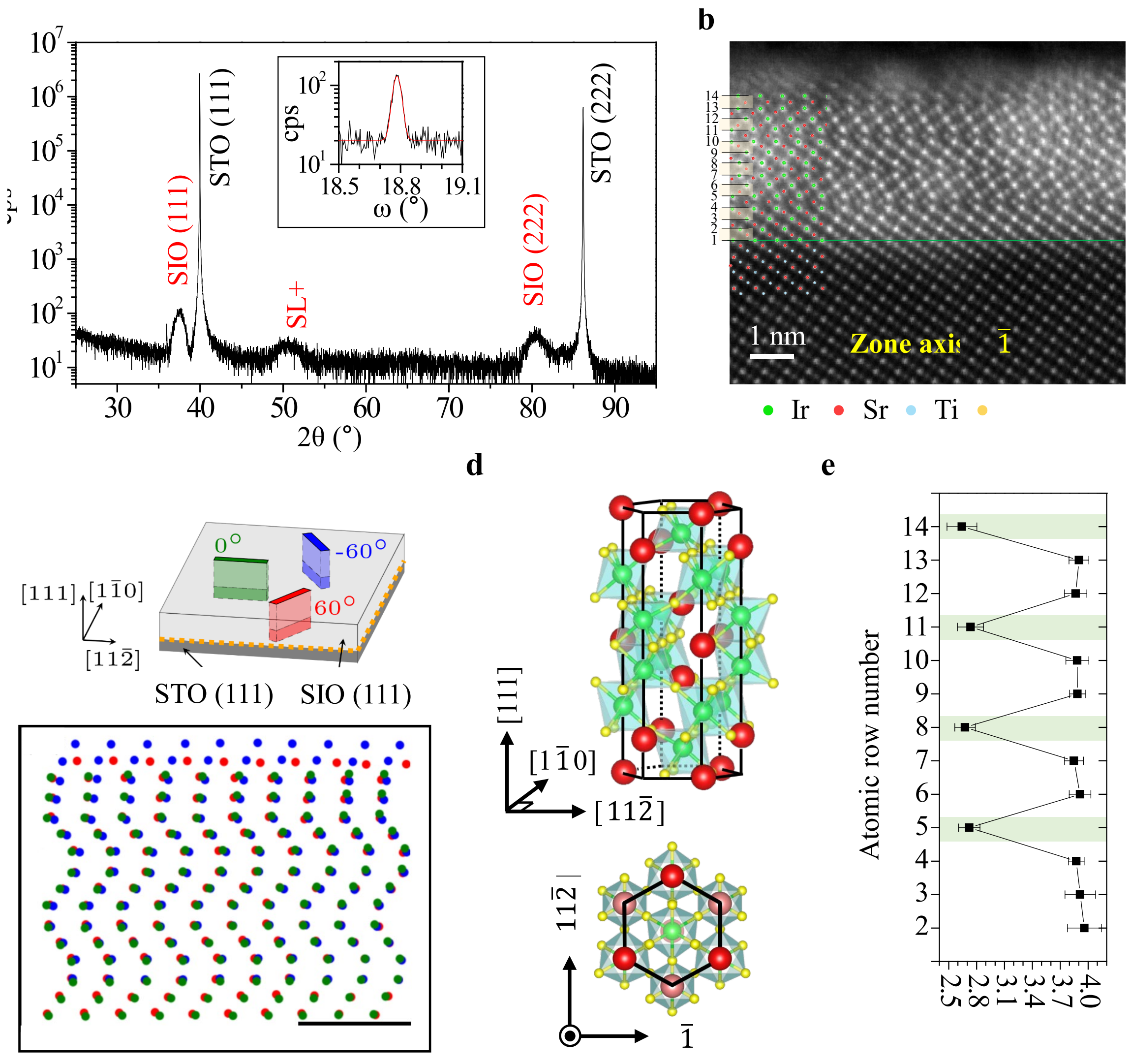

**Figure 1. Structural properties of SIO (111) films grown on STO (111) substrates**. **a**, X-ray diffractogram of a 2θ−ω scan. The inset shows a rocking curve of the SIO (111) diffraction peak. **b**, Scanning transmission electron microscopy image along the $[1\bar{1}0]$ zone axis. The numbers on the left denote the atomic rows in the image containing Ir atoms, counted from the interface. **c**, Arrangement of Ir atoms, generated from STEM images along three different zone axes, as illustrated in the schematic above. See Supporting information for details. **d**, Three-dimensional hexagonal crystal structure as inferred from the STEM data. The figure below shows a top view of the crystal structure. The lattice vectors are specified in relation to the STO substrate. **e**, The average Ir-Ir distance in the face-shared (green colored) and the corner-shared (uncolored) octahedra. The atomic row numbers $i$ refer to the numbers in **b** with the Ir-Ir distance given for each row being determined from the averaged distances of Ir atoms in this row and the next row $i$-1 to the interface.

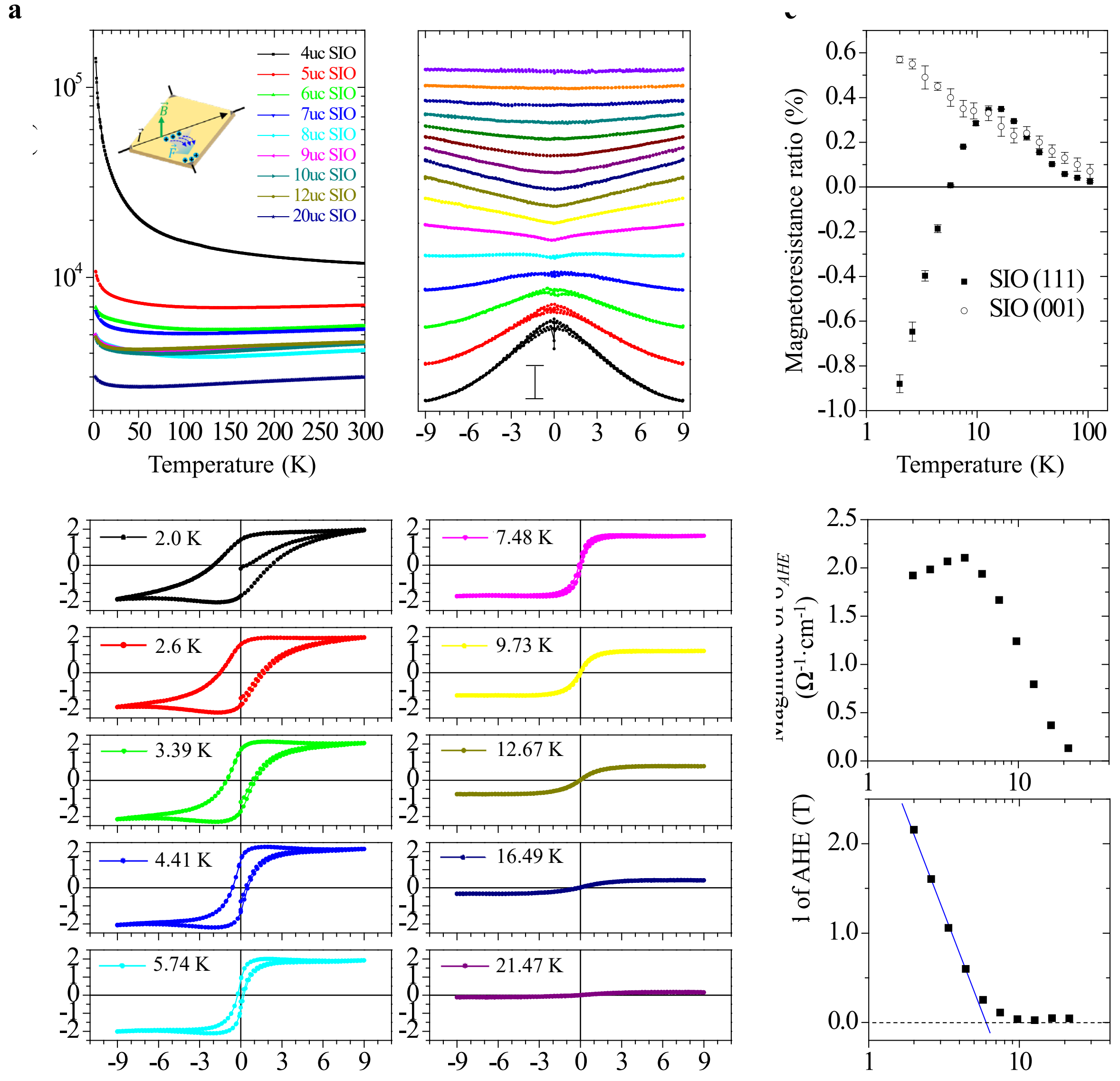

**Figure 2. Magnetotransport properties**. **a**, Temperature-dependent resistance for films with different thicknesses. A metal-insulator transition occurs at 5uc. **b**, MR of a 9 uc thick film. **c**, MR ratio of a SIO (001) (empty dots) and a SIO (111) (filled dots) film, with the MR ratio given as percentage of the average of $(R_{9T} - R_{0T})/R_{0T}$ and $(R_{-9T} - R_{0T})/R_{0T}$. The error bars are calculated from the difference of these two values. **d**, Anomalous Hall conductivity, calculated as $\sigma_{AHE} = \rho_{AHE}/(\rho_{AHE}{}^2 + \rho_{xx}{}^2)$, of a 9 uc thick film after subtracting the linear component for the ordinary Hall effect. Below about 30 K, the AHE appears and increases with decreasing temperature. Distinct hysteresis loops can be seen below 7 K. **e**, Magnitudes of the anomalous Hall conductivity, as obtained from the average of the saturated values $\sigma_{AHE}(6T)$ and $\sigma_{AHE}(-6T)$. **g**, Coercive field of the AHE as a function of temperature. Plotted is the average of the magnitudes of the magnetic field at which an anomalous Hall conductivity of zero is observed for each polarity.

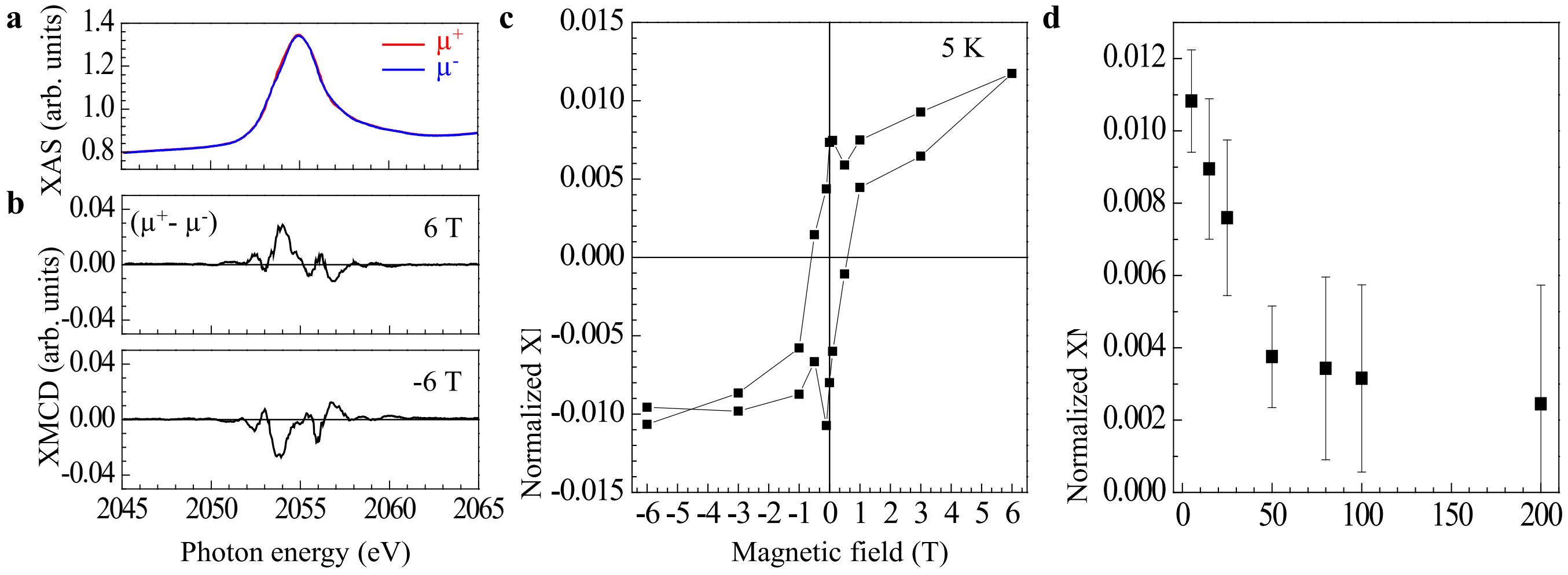

**Figure 3. Magnetic properties**. **a**, X-ray absorption spectroscopy (XAS) spectra at the Ir $M_5$ edge, measured at 5 K and 6 T with circularly right (red curve, $\mu^+$) and circularly left polarization (blue curve, $\mu^-$). **b**, X-ray circular dichroism (XMCD), $\mu^+ - \mu^-$, at 5 K and magnetic fields of 6 and -6 T. **c**, M-H hysteresis loop at 5 K. Each data point represents the integral over the corresponding XMCD signal. **d**, XMCD integral as a function of temperature, recorded in remanence after magnetization in a field of - 6T. The error bars are estimated from the scatter of repeated measurements.

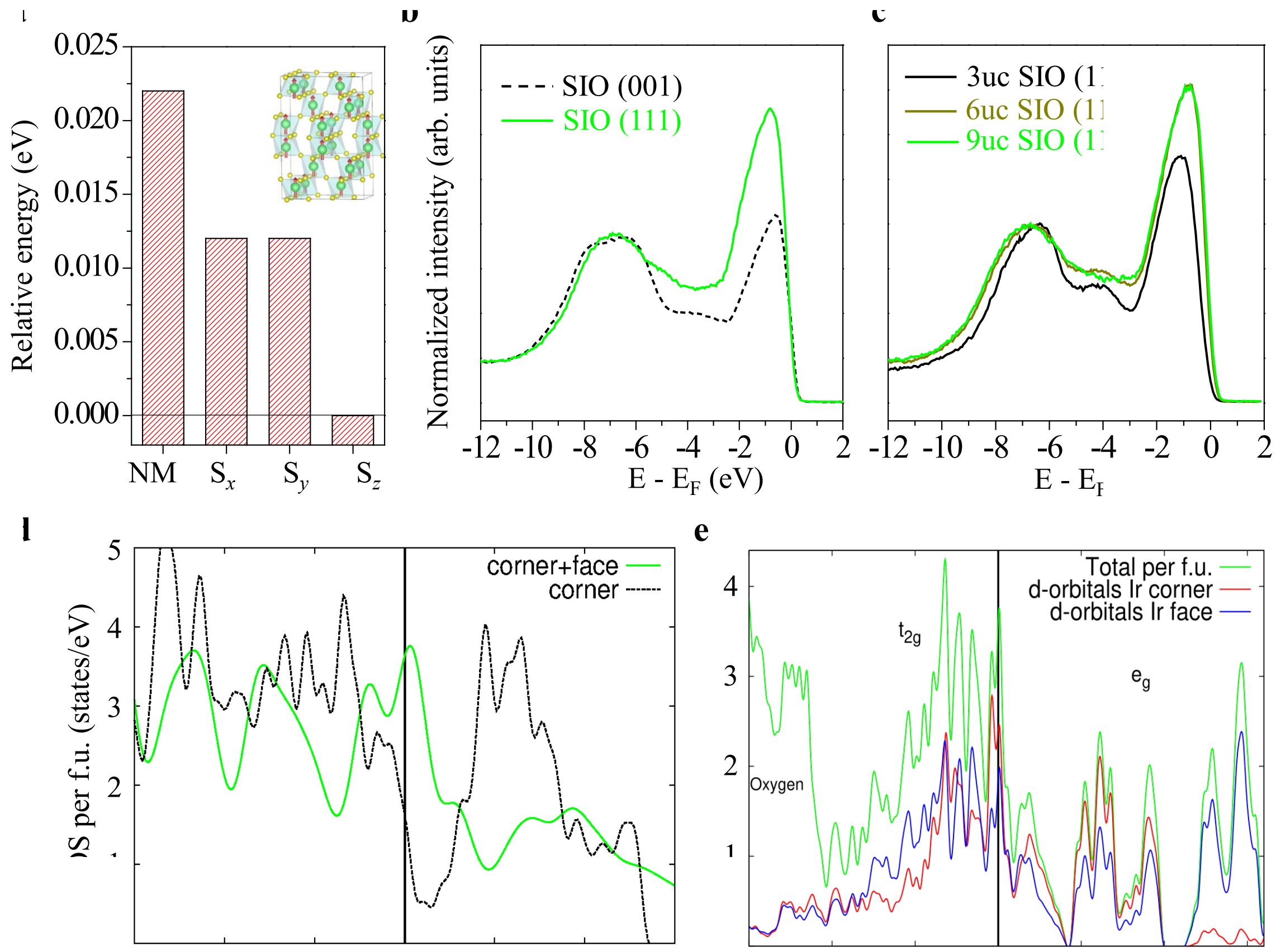

**Figure 4. Electronic structure of SIO (001) and (111) films from DFT + U and XPS**. **a**, Ground state energies of non-magnetic state (NM) and for spin orientations along x, y and z axes ($S_x$, $S_y$ and $S_z$). The energies are referred to the energy of the $S_z$ state. **b**, Angle-integrated valence band spectra of a SIO (001) film with a thickness of 10 uc (black dashed line) and a SIO (111) film with a thickness of 9 uc (green solid line) using X-ray photoelectron spectroscopy (Al Kα = 1486.6 eV). **c**, Angle-integrated valence band spectra of SIO (111) films with different thicknesses (3, 6 and 9 uc). **d**, DFT + U density of states for SIO (001) and (111) films, depicted by black dashed and solid green lines, rspectively. In the SIO (001) film, only corner-shared octahedra are present, while the SIO (111) film comprises a combination of corner- and face-shared octahedra. **e**, Total Ir 5d density of states and decomposition into partial density of states for Ir in corner- and face-shared octahedra.

## 4. Experimental Section/Methods

*Growth of SIO thin films and X-ray diffraction*: The pellets were prepared by stoichiometrically mixing $SrCO_3$ (99.9%) and $IrO_2$ (99.9%) powder (Sigma-Aldrich) and applying high pressure to obtain disk-shaped pellets with a diameter of 1 inch. The pellets then were calcinated at 850 ˚C for 12 hours under an oxygen flow of 0.5 l/min. After calcination, the pellets were grounded to a fine powder again, and pressed and calcinated as before to obtain the final targets for pulsed-laser deposition. To prepare STO substrates (CrysTec GmbH) with Ti termination, the substrates were successively cleaned with acetone, isopropanol and deionized water and then immersed in buffered hydrofluoric acid (1:6) for 30 seconds. The STO substrates then were heated at 1000 ˚C for 3 hours in an oxygen flow of 0.5 L per minute. SIO thin films were deposited on Ti-terminated STO (111) and (001) substrates by pulsed laser deposition using a KrF excimer laser ($\lambda$ = 248 nm). The substrate temperature during film growth was held at 600 ˚C (STO (111)) and 680 ˚C (STO (001)) in an oxygen environment of 0.08 mbar. Laser fluence and repetition rates were set to ~2 J $cm^{-2}$ and 3 Hz, respectively. All films were in-situ cooled down to room temperature at a rate of 50 ˚C $min^{-1}$ in an oxygen atmosphere of 0.08 mbar. The *c*-axis lattice parameters of the as-grown SIO thin films were characterized using a four-circle X-ray diffractometer (Bruker D8 Discovery) with Cu $K_{\alpha 1}$ radiation.

*X-ray photoelectron spectroscopy*: Angle-integrated valence band spectra of SIO (001) and (111) films were recorded *in situ* at room temperature by X-ray photoelectron spectroscopy with an Al K$\alpha$ source (1486.6 eV) (Scienta Omicron GmbH). The spectra were taken in normal emission geometry.

*Transmission electron microscopy (TEM)*: High-resolution scanning TEM (HR-STEM) was carried out using a TF Titan$^3$ 80–300 probe-corrected instrument (Thermo Fisher Scientific Inc., USA) operated at 300 kV acceleration voltage. Scan images were recorded with a high-angle annular dark-field (HAADF) detector at a camera length of 73 mm (collection angle 80–200 mrad), a probe semi-convergence angle of 21.5 mrad, a pixel size of 12.6 pm and a dwell time of 6 μs. In order to fit the atom positions of the HAADF-STEM images with pm-scale precision, the coarse positions were first determined using a spatial band-pass filter, i.e., the so-called difference of Gaussians (DoG) method [45]. Then, both the higher local contrast of the Ir atoms and their approximate distances were used to distinguish them from the other atomic species.

Finally, the precise Ir atom centers were obtained by means of two-dimensional Gaussian peak fitting.

*Electrical transport measurement*: Magnetotransport measurements were carried out using a Quantum Design Physical Property Measurement System (PPMS-9T) with resistivity option, by applying an AC current of 10 μA. As-grown samples (5 mm x 5 mm) were mounted on PPMS-sample pucks. Electrical connections in van-der-Pauw geometry were made with Al wires that were bonded directly to the sample surface and to the metal pads of the sample pucks using an ultrasonic wire bonder. A magnetic field was applied perpendicular to the sample surface. Magnetoresistance and Hall measurements were performed from the lowest temperature of 2 K up to 300 K in 20 steps of a logarithmic temperature increase.

*XAS and XMCD measurements*: X-ray absorption spectroscopy (XAS) and X-ray magnetic circular dichroism (XMCD) measurements at the Ir $M_5$ edge were carried out at the DEIMOS beamline at the SOLEIL synchrotron facility (France) [40]. All samples were transferred into an ultrahigh vacuum suitcase (base pressure $\sim 10^{-10}$ mbar) to prevent degradation in air. As the signal, the total electron yield (TEY) was measured with a drain current detector, while the incident X-ray intensity was monitored by the photocurrent of a gold mesh with 25% transmission. A large beam spot size of 800 × 800 $\mu m^2$ on the sample was used to avoid radiation damage. All samples were measured in a single sequence with alternating polarizations ($\mu^+\mu^+\mu^-\mu^-\mu^+$), with the initial $\mu^+$ spectrum being excluded from the XMCD analysis. To isolate the XMCD signal, a Shirley background was subtracted to account for the absorption into the free-electron continuum. (Supporting Information, Fig. S5)

*Calculation of electronic band structure*: DFT calculations were performed using the VASP ab initio simulation package [46], employing the projector-augmented wave (PAW) method [47] within the generalized gradient approximation (GGA). The revised Perdew-Burke-Ernzerhof (PBEsol) [48] functional was used, both with and without the inclusion of spin-orbit coupling (SOC). The plane-wave cutoff energy for expanding the Kohn-Sham orbitals was set to 400 eV. Brillouin-zone integration was carried out using a Monkhorst-Pack [49] k-point mesh of 10 × 6 × 4, centered at the M-point. On-site Hubbard Coulomb repulsion and Hund's exchange were incorporated using the DFT+U method with realistic values which best fit the experimental data of bulk orthorhombic $SrIrO_3$. These electronic parameters were determined to be U = 0.8 eV and $J_H = 0.15U$ [50].

**Author Contributions**

**All authors discussed the results and contributed to the completion of the manuscript. J.S.L. and C.A. wrote the manuscript. Thin film growth was performed by J.S.L. and M.S. Magnetotransport, XPS, and XRD measurements were carried out by J.S.L. XMCD and XAS experiments were conducted by J.S.L. with the assistance of F.C. and P.O. at the SOLEIL-DEIMOS beamline. Theoretical calculations were performed by C.A., A.F., M.C., and G.S. TEM measurements and analyses were conducted by M.K., P.P., D.W., S.P., J.S., and A.L. Angular magnetotransport under high magnetic fields was investigated by N.P., B.M., and L.V. B.B. supervised the TEM and high-field magnetotransport studies. M.S. and R.C. supervised and coordinated the overall project.**

**Acknowledgements**

**J. S. L. and C. A. contributed equally to this work. We gratefully acknowledge funding by the Deutsche Forschungsgemeinschaft (DFG) through Projects No. 461150024 and No. 431448015, through the Collaborative Research Centers SFB 1143 (project C04, project id 247310070) and SFB 1170 (project C08, project id 258499086) as well as the Würzburg–Dresden Cluster of Excellence *ct.qmat*, EXC 2147 (project id 390858490). We also appreciate financial support by the Bundesministerium für Bildung und Forschung (BMBF) through the project OperandoHAXPES (project number 05K22WW1). L.V. acknowledges support from the french ANR under projects ANR-24-CE92-0021-01 and ANR-23-CPJ1-0158-01, and from the Leibniz Association through the Leibniz Competition. C. A. is supported by the Foundation for Polish Science project "MagTop" No. FENG.02.01-IP.05-0028/23 co-financed by the European Union from the funds of Priority 2 of the European Funds for a Smart EconomyProgram2021–2027 (FENG). C.A. and M.C. acknowledge support from PNRR MUR project PE0000023-NQSTI. A.F. was supported by the Polish National Science Center under Project No. 2020/37/B/ST5/02299. We acknowledge the computing resources and support provided by the computer systems of the Interdisciplinary Centre for Mathematical and Computational Modelling at the University of Warsaw, through Grants No. g911418, No. g91-1419, No. g91-1426, No. g96-1808, and No. g96-1809. XMCD and XAS were performed at the DEIMOS beamline at the SOLEIL Synchrotron facility, France (proposal number: 20230230). We are grateful to the SOLEIL staff for their excellent support. We also gratefully acknowledge Ivan Soldatov and Rudolf Schäfer at the Leibniz Institute for Solid State and Materials**

**Research Dresden for valuable discussions regarding the magnetic domains of the SIO (111) films.**

## References

[1] S. Bhatti, R. Sbiaa, A. Hirohata, H. Ohno, S. Fukami and S. N. Piramanayagam, *Material Toady* **2017**, *20*, 530−548.

[2] D. D. Awschalom and M. E. Flatte, *Nat. Phys.* **2007**, *3*, 153−159.

[3] S. D. Bader and S. S. P. Parkin, *Annu. Rev. Condens. Matter Phys.* **2010**, *1*, 71–88.

[4] Q. Wang, Y. Gu, C. Chen, F. Pan and C. Song, *J. Phys. Chem. Lett.* **2022**, *13*, 10065−10075.

[5] F. Trier, P. Noël, J.-V. Kim, J.-P. Attane, L. Vila and M. Bibes, *Nat. Rev. Mater.* **2022**, *7*, 258−274.

[6] L. Liu, Q. Qin, W. Lin, C. Li, Q. Xie, S. He, X. Shu, C. Zhou, Z. Lim, J. Yu, W. Lu, M. Li, X. Yan, S. J. Pennycook and J. Chen, *Nat. Nanotechnol.* **2019**, *14*, 939−944.

[7] S. Hu, D.-F. Shao, H. Yang, C. Pan, Z. Fu, M. Tang, Y. Yang, W. Fan, S. Zhou, E. Y. Tsymbal and X. Qiu, *Nat. Commun.* **2022**, *13*, 4447.

[8] K.-W. Kim, B.-G. Park and K.-J. Lee, *npj Spintronics* **2024**, 2:8.

[9] A. Davidson, V. P. Amin, W. S. Aljuaid, P. M. Haney and X. Fan, *Phys. Lett. A* **2020**, *384*, 126228.

[10] J. Sinova, S. O. Valenzuela, C. H. Back and T. Jungwirth, *Rev. Mod. Phys.* **2015**, *87*, 1213−1259.

[11] Y. Hibino, T. Taniguchi, K. Yakushiji, A. Fukushima, H. Kubota and S. Yuasa, *Nat. Commun.* **2021**, *12*, 6254.

[12] M. Aoki, Y. Yin, S. Granville, Y. Zhang, N. V. Medhekar, L. Leiva, R. Ohshima, Y. Ando and M. Shiraishi, *Nano Lett.* **2023**, *23*, 6951−6957.

[13] L. Liu, C. Zhou, X. Shu, C. Li, T. Zhao, W. Lin, J. Deng, Q, Xie, S. Chen, J. Zhou, R. Guo, H. Wang, J. Yu, S. Shi, P. Yang, S. Pennycook, A. Manchon and J. Chen, *Nat. Nanotechnol*, **2021**, *16*, 277−282.

[14] Y. J. Chang, C. H. Kim, S.-H. Phark, Y. S. Kim, J. Yu and T. W. Noh, *Phys. Rev. Lett.* **2009**, *103*, 057201.

[15] M. Kim and B. I. Kim, *Phys. Rev. B* **2015**, *91*, 205116.

[16] Y. Jo, Y. Kim, S. Kim, E. Ryoo, G. Noh, G.-J. Han, J. H. Lee, W. J. Cho, G.-H. Lee, S.-Y. Choi and D. Lee, *Nano Lett.* **2024**, *24*, 7100−7107.

[17] A. Biswas, K.-S. Kim and Y. H. Jeong, *J. Appl. Phys*. **2014**, *116*, 213704.

[18] D. J. Groenendijk, C. Autieri, J. Girovsky, M. Carmen Martinez-Velarte, N. Manca, G. Mattoni, A. M. R. V. L. Monteiro, N. Gauquelin, J. Verbeeck, A. F. Otte, M. Gabay, S.

Picozzi and A. D. Caviglia, *Phys. Rev. Lett.* **2017**, *119*, 256403.

[19] P. Schütz, D. D. Sante, L. Dudy, J. Gabel, M. Stübinger, M. Kamp, Y. Huang, M. Capone, M.-A. Husanu, V. N. Strocov, G. Sangiovanni, M. Sing and R. Claessen, *Phys. Rev. Lett.* **2017**, *119*, 256404.

[20] J. Matsuno, K. Ihara, S. Yamamura, H. Wadati, K. Ishii, V. V. Shankar, H.-Y. Kee and H. Takagi, *Phys. Rev. Lett.* **2015**, *114*, 247209.

[21] J. Liu, X. Zhang, Y. Ji, X. Gao, J. Wu, M. Zhang, L. Li, X. Liu, W. Yan, T. Yao, Y. Yin, L. Wang, H. Guo, G. Cheng, Z. Wang, P. Gao, Y. Wang, K. Chen and Z. Liao, *J. Phys. Chem. Lett.* **2022**, *13*, 11946.

[22] G. R. Jeffrey, K.-H. L. Eric and K. Hae-Young, *Annu. Rev. Condens. Matter Phys.* **2016**, *7*, 195.

[23] T. C. van Thiel, J. Fowlie, C. Autieri, N. Manca, M. Šiškins, D. Afanasiev, S. Gariglio and A. D. Caviglia, *ACS Mater. Lett.* **2020**, *2*, 389.

[24] G. Cao, V. Durairaj, S. Chikara, L. E. DeLong, S. Parkin, and P. Schlottmann, *Phys. Rev. B* **2007**, *76*, 100402(R).

[25] S. Okamoto and D. Xiao, *J. Phys. Soc. Jpn.* **2018**, *87*, 041006.

[26] T. J. Anderson, S. Ryu, H. Zhou, L. Xie, J. P. Podkaminer, Y. Ma, J. Irwin, X. Q. Pan, M. S. Rzchowski and C. B. Eom, *Appl. Phys. Lett.* **2016**, *108*, 151604.

[27] X. Zheng, S. Kong, J. Zhu, J. Feng, Z. Lu, H. Du, B. Ge, T. Wu, Z. Wang, M. Radovic and R.-W. Li, *Thin Solid Films* **2020**, 709, 138119.

[28] W. Zhao, M. Gu, D. Xiao, Q. Li, X. Liu, K. Jin, H. Guo, M. Meng and J. G uo, *Phys. Rev. Materials* **2024**, *8*, 105001.

[29] H. Shilin, L. Meilin, X. Wen, Y, Zhan, L. Junhua, H. Yuhao, D. Zhixiong, G. Xiaofei, S. Ziyue, W. Lei, Y. Runze, Z. Qinghua, G. Yulin, C. Kai and L. Zhaoliang, *Adv. Funct. Mater.* **2025**, *35*, 2414425.

[30] J. M. D. Coey, *Magnetism and Magnetic Materials* (Cambridge Univ. Press, **2009**).

[31] F. Matsukura, H. Ohno, A. Shen and Y. Sugawara, *Phys. Rev. B* **1998**, *57*, R2037−R2040.

[32] K. Yasuda, A. Tsukazaki, R. Yoshimi, K. S. Takahashi, M. Kawasaki and Y. Tokura, *Phys. Rev. Lett.* **2019**, *117*, 127202.

[33] K. Jaiswal, A. G. Zaitsev, R. Singh, R. Schneider and D. Fuchs, *AIP Advances* **2019**, *9*, 125034.

[34] A. Biswas, K.-S. Kim and Y. H. Jeong, *J. Appl. Phys*. **2014**, *116*, 213704.

[35] F.-X. Wu, J. Zhou, L. Y. Zhang, Y. B. Chen, S.-T. Zhang, Z.-B. Gu, S.-H. Yao and Y.-F. Chen, *J. Phys.: Condens. Matter* **2013**, *25*, 125604.

[36] K. Ueda, R. Kaneko, H. Ishizuka, J. Fujioka, N. Nagaosa and Y. Tokura, *Nat. Commun.* **2018**, *9*, 3032.

[37] W. J. Kim, T. Oh, J. Song, E. K. Ko, Y. Li, J. Mun, B. Kim, J. Son, Z. Yang, Y. Kohama, M. Kim, B.-J. Yang and T. W. Noh, *Sci. Adv.* **2020**, *6*, eabb1539.

[38] G. Vértesy, I. Tomáš, L. Půst and J. Pačes, *J. Appl. Phys.* **1992**, *71*, 3462–3466.

[39] N. Nagaosa, J. Sinova, S. Onoda, A. H. MacDonald and N. P. Ong, *Rev. Mod. Phys.* **2010**, *82*, 1539–1592.

[40] P. Ohresser, E. Otero, F. Choueikani, K. Chen, S. Stanescu, F. Deschamps, T. Moreno, F. Polack, B. Lagarde, J.-P. Daguerre, F. Marteau, F. Scheurer, L. Joly, J.-P. Kappler, B. Muller, O. Bunau, Ph. Sainctavit, *Rev. Sci. Instrum.* **2014**, *85*, 013106.

[41] S. Sen Gupta, J. A. Bradley, M. W. Haverkort, G. T. Seidler, A. Tanaka, and G. A. Sawatzky, *Phys. Rev. B* **2011**, *84*, 075134.

[42] S. M. Butorin, *arXiv* **2025**, *arXiv:2505.14284*.

[43] C.-K. Duan, M. F. Reid, and G. Ruan, *Curr. Appl. Phys.* **2006**, *6*, 359–362.

[44] M. Ye, H.-S. Kim, J.-W. Kim, C.-J. Won, K. Haule, D. Vanderbilt, S.-W. Cheong, G. Blumberg, *Phys. Rev. B* **2018**, 98, 201105(R).

[45] F. Akram, M. A. Garcia and D. Puig, *Sci. Rep.* **2017**, *7*, 14984.

[46] G. Kresse and D. Joubert, *Phys. Rev. B* **1999**, *59*, 1758.

[47] P. E. Blöchl, *Phys. Rev. B* **1994**, *50*, 17953.

[48] J. P. Perdew, A. Ruzsinszky, G. I. Csonka, O. A. Vydrov, G. E. Scuseria, L. A. C onstantin, X. Zhou, and K. Burke, *Phys. Rev. Lett.* **2008**, *100*, 136406.

[49] J. D. Pack and H. J. Monkhorst, *Phys. Rev. B* **1977**, *16*, 1748.

[50] N. Manca, D. J. Groenendijk, I. Pallecchi, C. Autieri, Lucas MK Tang, F. Telesio, G. Mattoni, A. McCollam, S. Picozzi, and A. D. Caviglia. *Phys. Rev. B* **2018**, *97*, 081105.

Supporting Information

## S1. RHEED monitoring

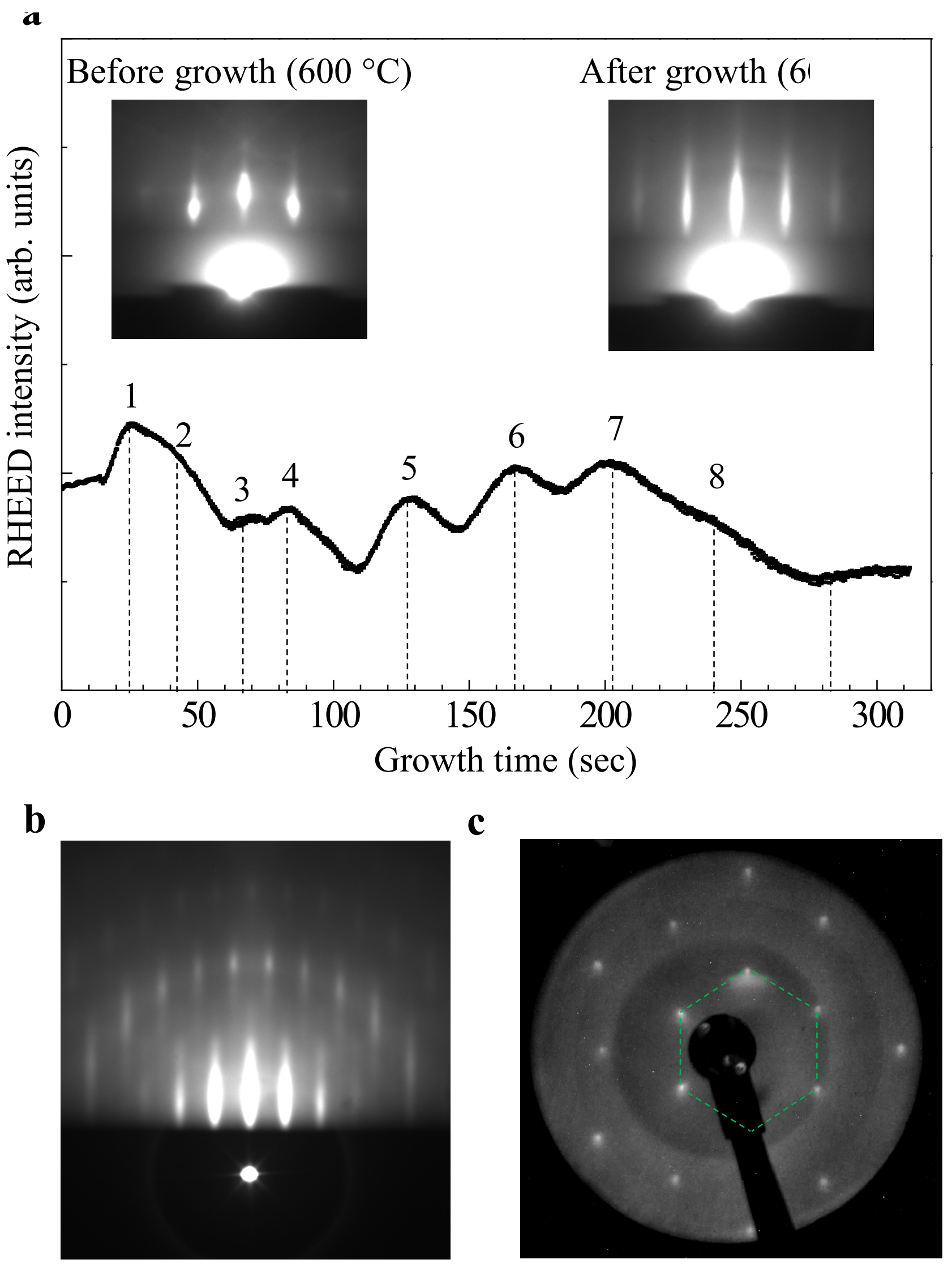

**Figure S1. a,** RHEED oscillations during growth of a 9 uc thick $SrIrO_3$ (111)/$SrTiO_3$ (111) film. The inset figures show RHEED patterns just before and after $SrIrO_3$ (SIO) film growth. **b**, RHEED pattern of a SIO film (9 uc) at room temperature. **c**, LEED image of a SIO film (9 uc) at room temperature. The green dashed lines indicate the hexagonal structure.

## S2. Atomic arrangement from TEM images

The atomic arrangement of our SIO (111) films is shown in Fig. S2a. The green dots represent the positions of Ir atoms, while the red dots mark the positions of Sr atoms. In Fig. S2b, in left panel, the mean value of the *c*-axis Ir-Ir spacing (marked as *c* in Fig. S2a) is displayed, where the averaging is carried out over all atomic rows of the TEM image in Fig. S2c. As can be seen, an Ir-Ir spacing of about 2.2 Å within the 3uc-thick stacks contrasts with a distance of about 2.7 Å at the interfaces. Notably, the interface region exhibits face-shared octahedra, whereas the interior of the stacks consists of corner-shared octahedra, characteristic of a perovskite structure. Additionally, the average Ir-Ir in-plane distance amounts to about 4.8 Å (Fig. S2b, right panel), a value nearly identical to the Ti-Ti spacing of the STO (111) substrate. The TEM image taken along the $[11\bar{2}]$ zone axis provides further confirmation of the epitaxial strain between the SIO film and the STO substrate (Fig. S2c). The white dots in the image represent Ir atoms, and it is clearly seen that every 3 unit cells along the z-axis, the interfaces exhibit a larger spacing (as shown in Fig. S2c). The schematic on the right side of Fig. S2c visualizes the atomic arrangement along the $[11\bar{2}]$ zone axis. To explore the structural symmetry of the SIO (111) film, we obtained TEM images from three other lamellae, cut at different angles of 0º, -60º, and 60º with respect to the $[11\bar{2}]$ axis (Fig. S2d). The colored dots mark Ir atomic positions. Since the Ir atomic positions for different angles can nicely be overlaid, a sixfold symmetric structure is confirmed.

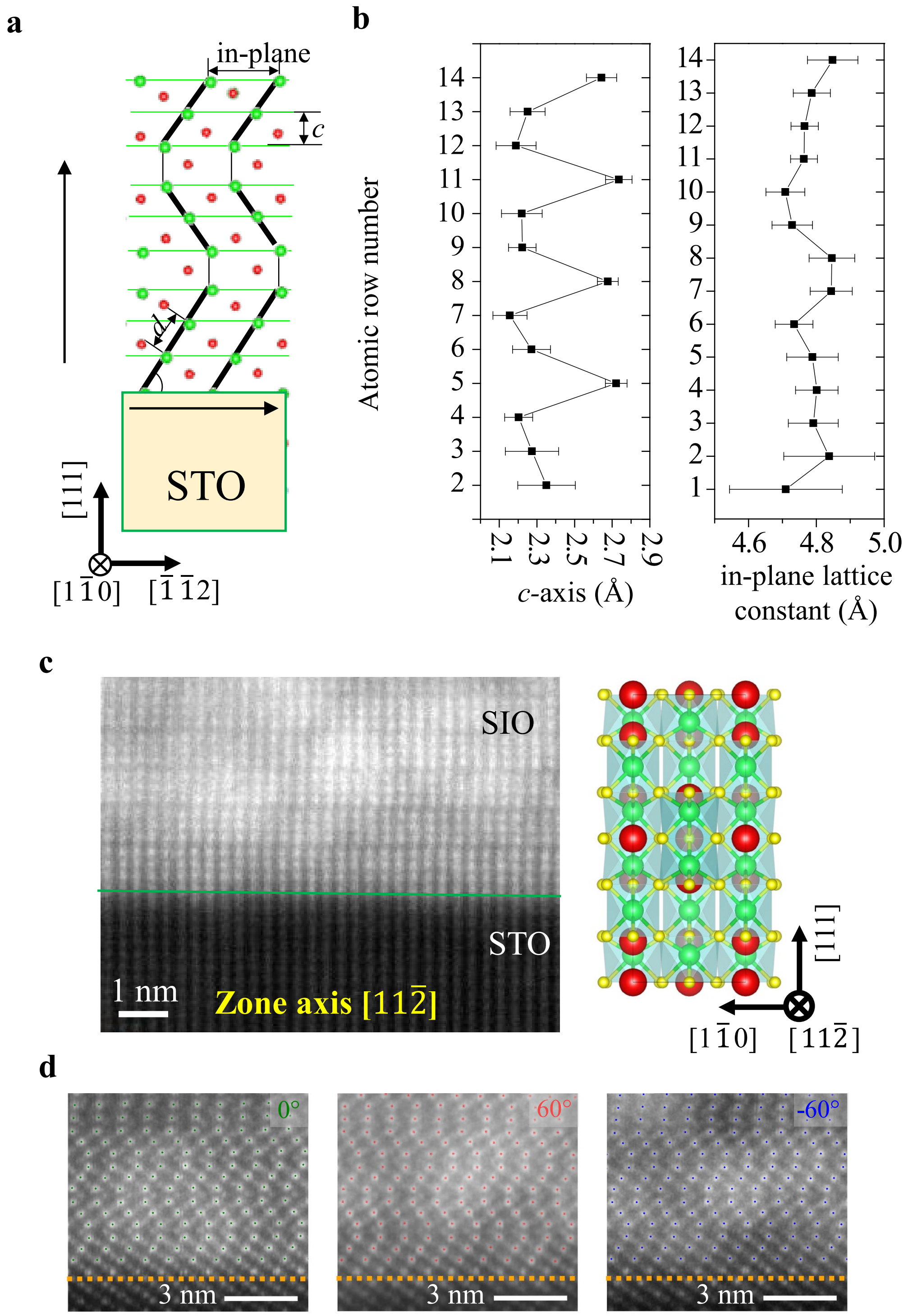

**Figure S2. a,** Schematic of atomic arrangement as derived from the TEM image in Fig. 1b of the manuscript and **b**, corresponding averaged Ir-Ir atomic spacing between row number *i* and the adjacent row *i-1* along the out-of-plane direction (left), and the averaged Ir-Ir atomic spacing within each row *i* along the in-plane direction (right). **c**, TEM image of an SIO (111) film along the [11$\bar{2}$] zone axis and corresponding atomic structure. **d**, TEM images used to derive the atomic arrangement. The angles displayed in the top right corner of each image refer to the angle of the corresponding lamellae with respect to the [1$\bar{1}$0] zone axis.

## S3. Magnetoresistance of SIO (001) films

To understand how film orientation influences the electrical properties, we measured the temperature-dependent resistance of the 10 uc SIO (001) and 9 uc SIO (111) films. The SIO (001) film shows a resistance behavior similar to that previously reported [1]. The resistivity at higher temperatures, up to room temperature, follows a power-law with an exponent less than 2 [1]. The low temperature regime hints to weak localization. The 9 uc SIO (111) film shows metallic behavior down to about 40 K and then a pronounced upturn (Fig. S3a). The magnetoresistance of the SIO (001) film with a field up to ± 9 T is always positive over the whole measured temperature range. The positive MR is likely attributed to weak antilocalization induced by strong spin-orbit coupling [2].

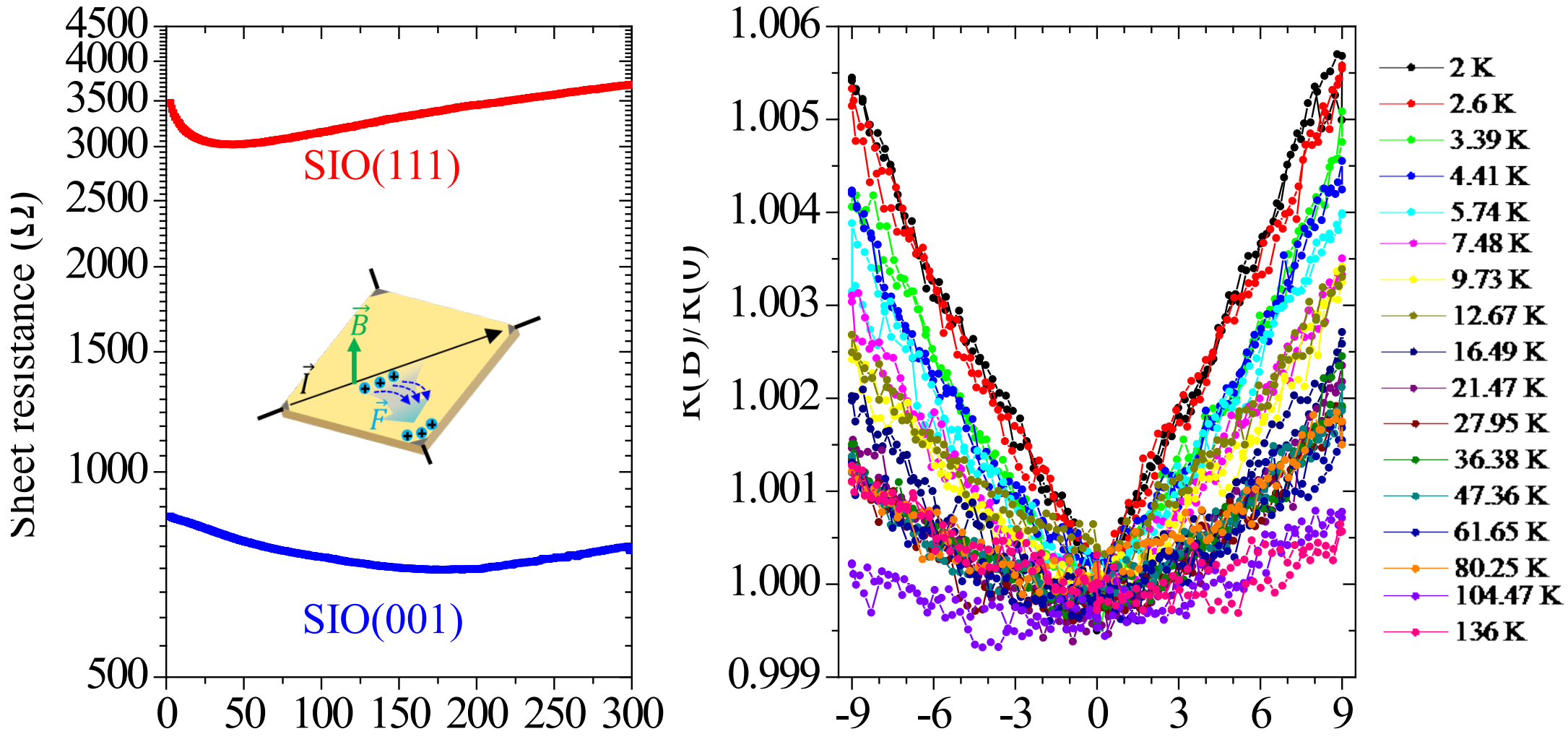

**Figure S3. a**, Sheet resistance of SIO (001) and (111) films as a function of temperature. The inset is a schematic of the van-der-Pauw method used. **b**, Magnetoresistance of the SIO (001) film for different temperatures.

## S4. Hall measurements of SIO (001) and (111) films

The SIO (001) film exhibits a linear Hall resistance with a negative slope over the whole measured temperature range (Fig. S4a, left panel). The electron-dominant transport in SIO (001) was explained by the specific Fermi surface that features electron pockets with nearly linear band crossings and high Fermi velocities, as well as hole pockets with a large effective mass [3]. On the other hand, the SIO (111) film shows a Hall resistance with a positive slope over the whole measured temperature range and, surprisingly, exhibits anomalous Hall resistance in the low-temperature range (< 30 K) (Fig. S4a, right panel). The transport in the SIO (111) film is dominated by holes. The The Hall resistance of SIO (111) is composed of linear and hysteresis terms, which correspond to the ordinary and anomalous Hall components. The anomalous Hall resistance saturates at high magnetic fields. From the linear terms in both films, we can estimate the carrier density (Fig. S4b) and mobility (Fig. S4c). The carrier density and mobility of the SIO (111) film (red dots) saturate at about 30 K. This temperature corresponds to the magnetic transition in the SIO (111) film.

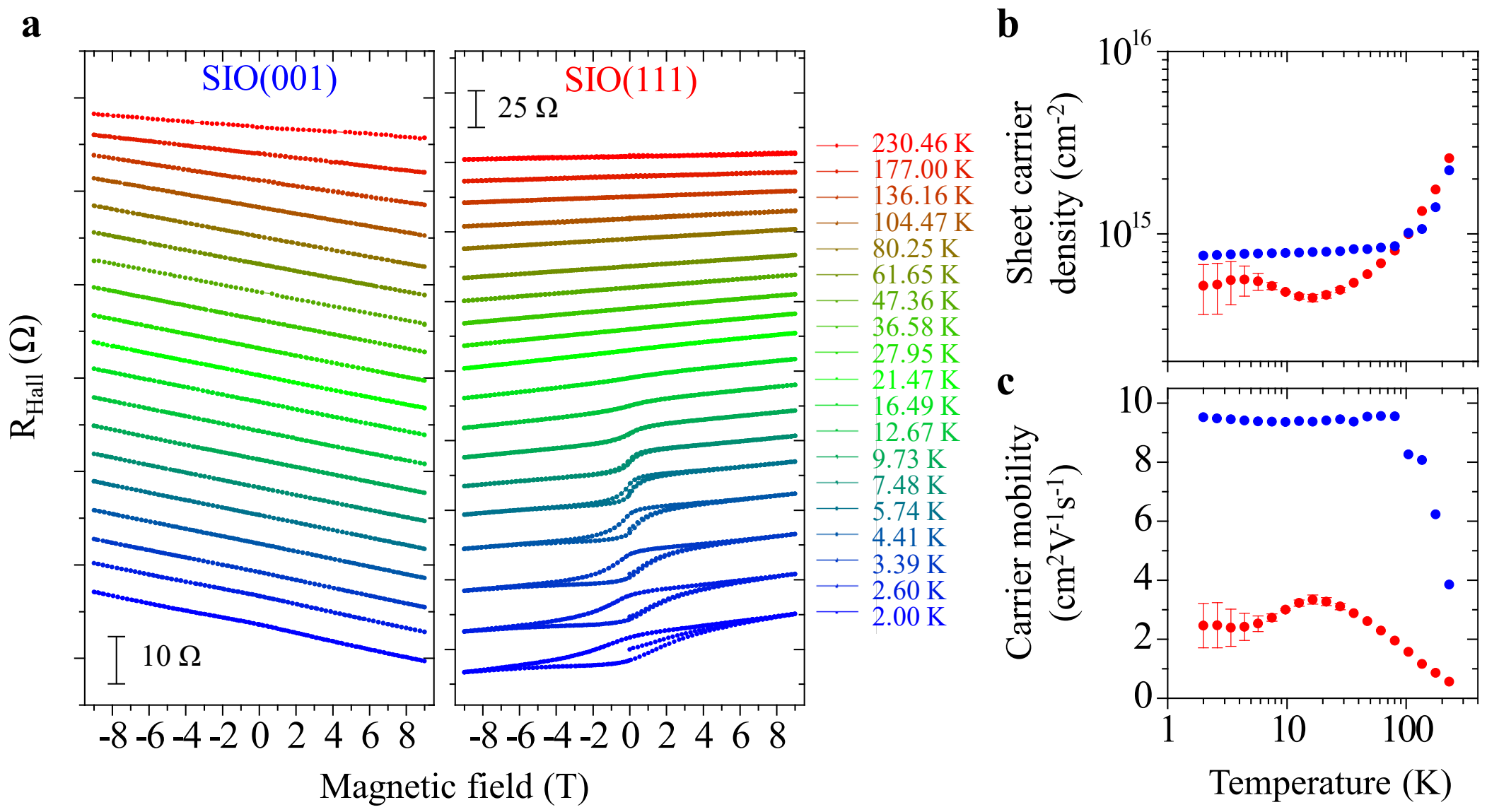

**Figure S4. a**, Hall resistances of SIO (001) and (111) films. Both are measured using van- der-Pauw geometry. The scales indicated correspond to a resistance of 10 Ω and 25 Ω under the application of an external current of 10 µA for the (001) and (111) films, respectively. Negative (positive) slopes of the Hall resistance signal electron- (hole)-dominated transport. **b**, The sheet carrier densities of the films are calculated from the linear slope of the Hall resistance at each temperature. In the region of the anomalous Hall effect in the SIO (111) film, the linear slopes at high magnetic fields (starting above the onset of the hysteresis loops) are used to evaluate the carrier densities. **c**, Carrier mobilities for the SIO films.

## S5 Anisotropic magnetotransport of SIO (111) films in high magnetic fields

To investigate the anisotropic magnetotransport properties of the SIO (111) film, Hallbar structures were fabricated using laser lithography followed by Ar ion etching. The Hallbars were aligned along the crystallographic orientations of STO substrate along [1 -1 0] and [1 1 -2] directions, with dimensions 70x10μm$^2$. As both Hallbars show qualitatively similar results, we present below the results of the Hallbar oriented along [1 -1 0]. The measurements were carried out at 2K on a 14T DynaCool PPMS with rotator, allowing the magnetic field could be rotated from the out-of-plane toward the in-plane configuration.

In Fig. S5a, we compare the magnetoresistance and transverse resistance data acquired at 2K on patterned Hall bars up to 14T to those presented in the main text (acquired up to 9T in van-der-Pauw geometry). The Hall effect observed quantitatively agrees with that observed on unpatterned films, with a typical 2D carrier density of $4.5 \times 10^{14}\ \mathrm{cm}^{-2}$ . The magnetoresistance is very small, about 0.1% at most, and the average mobility extracted is about $4\ \mathrm{cm}^2/\mathrm{V.s}$. Discrepancies between the unpatterned and patterned thin films could be due to local inhomogeneities or defects induced by the fabrication process. Above ~7.5T, an upturn in the magnetoresistance is observed. At the same field, the hysteresis loop of the Hall effect closes. This indicates that the negative magnetoresistance, observed at low magnetic field, is probably induced by magnetism.

Next, we study the Hall effect under a tilted magnetic field (Fig. S5b). To analyze only the hysteretic field-antisymmetric component, the Hall data were corrected to account for a longitudinal component in the signal. A background corresponding to the field-symmetric component, averaged between up and down sweeps:

$$R_{sym}^{av} = (R_{sym}^{up} + R_{sym}^{down})/2$$

with

$$R_{sym}^{up-down} = [R^{up-down}(B) + R^{up-down}(-B)]/2$$

was removed. This allows to remove the field-symmetric component of the transverse resistance, without removing the hysteretic part of the Hall effect. To obtain the anomalous Hall effect (AHE), a linear normal Hall component was removed. The linear normal Hall effect slope A was first fitted in the transverse resistance under a perpendicular field above 10T, and a background $A * \cos(\theta) * B$ was subsequently removed from the data at each tilt angle θ. The measurements were carried out with increasing $\theta$, from $\theta = 0°$ to $\theta = 90°$. We observe a butterfly-shaped AHE under a tilted field, with vanishing AHE at zero field for in-plane orientation ($\theta = 90°$). This is different from the expected AHE in a ferromagnet with out-of-plane anisotropy, for which the remanent AHE is expected to stay constant with tilt angle, while the saturated AHE decreases toward zero as the field is tilted toward in-plane orientation [4]. From the observed loops under titled field and the zero remanent AHE, we expect that saturation has not been reached and that the saturation field for orientations other than $\theta = 0°$ is above the maximum experimentally available field of 14T. Still, the presence of the ferromagnetic-like saturated hysteresis loop of the AHE in perpendicular field indicates the presence of a ferromagnetic coupling with out-of-plane anisotropy, as indicated by the XMCD and MOKE measurements.

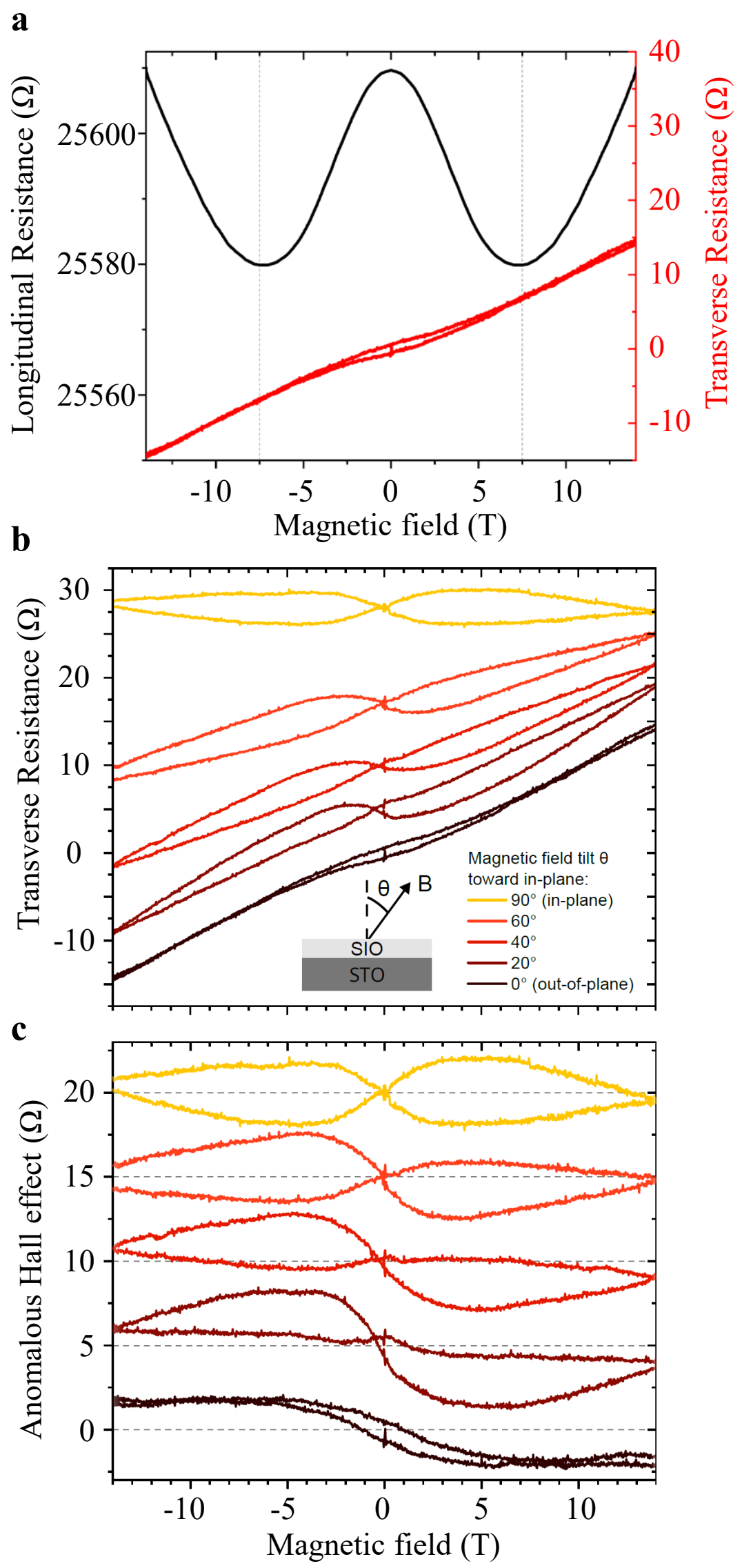

**Figure S5. a**, Longitudinal (black) and transverse (red) resistances from Hallbars patterned in an 8 uc SIO (111) film, under magnetic fields up to 14T. The upturn field is highlighted by grey dashed lines. **b**, Transverse resistance versus applied magnetic field, for different orientation of the field between out-of-plane (0°) toward in-plane (90°) configurations. Data is vertically shifted for clarity. **c**, AHE after removal of the normal Hall effect component. Data is vertically shifted for clarity. The dotted lines identify the corresponding zero of the AHE.

## S6. Magnetic field dependent XMCD

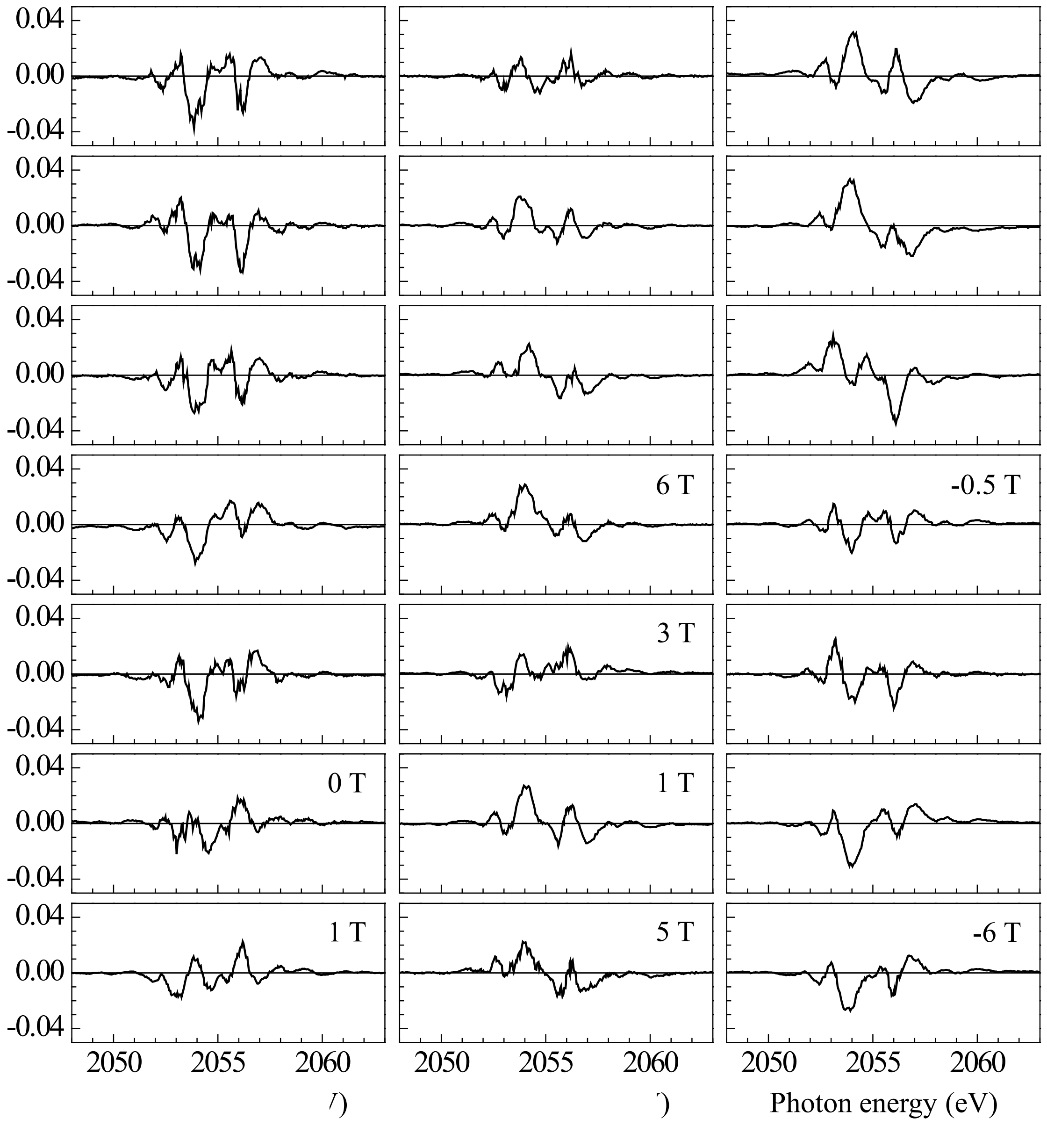

**Figure S6.** XMCD (X-ray Magnetic Circular Dichroism) of the Ir $M_5$ edge measured at 5K for varying magnetic field perpendicular to the sample surface, used to extract a M-H hysteresis loop. The magnetic field was cycled through the sequence -6 T, 0 T, 6 T, 0 T, and -6 T.

## S7. Temperature-dependent XAS and XMCD

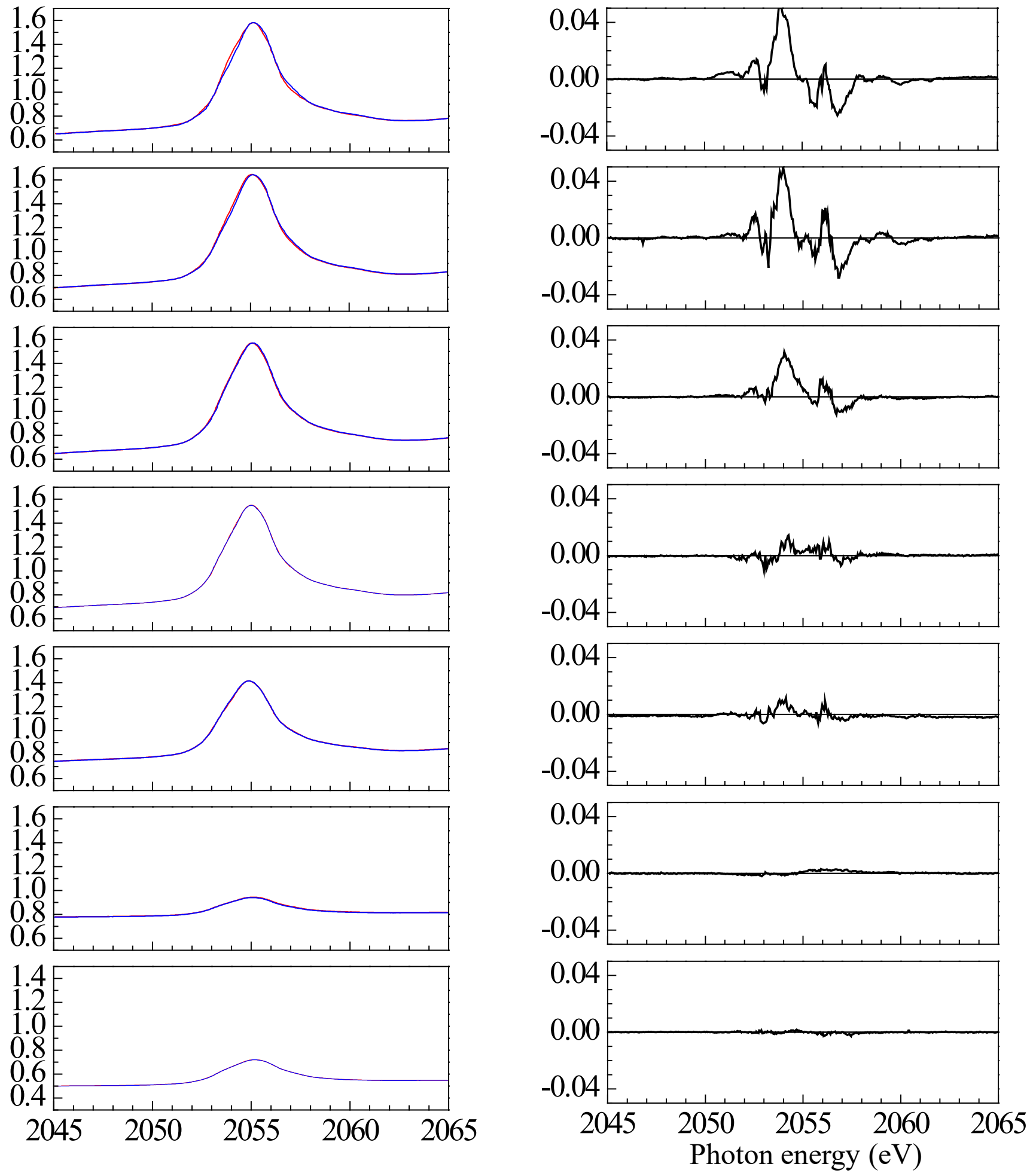

**Figure S7.** A pronounced decrease in the Ir $M_5$ absorption edge intensity is observed in the temperature-dependent XAS spectra of a SIO/STO (111) film. To track changes in the XMCD signal, it is normalized to the integrated XAS spectral weight. The origin of the intensity loss remains to be clarified, but it may be linked to electronic structure modifications in the SIO film induced by the cubic-to-tetragonal phase transition of the STO substrate at 105 K [5]. This transition involves oxygen octahedral rotations, leading to in-plane lattice contraction and elongation along the c-axis [6]. In SIO/STO (001) films, a slight increase in conductivity below 105 K suggests that the substrate transition affects octahedral rotations in the film, thereby influencing its electronic properties [7]. Although the SIO (111) orientation differs structurally, similar effects from the STO transition may also influence the 4f hole count in this case.

.

## S8. Calculation of magnetic moment as function of on-site Coulomb repulsion

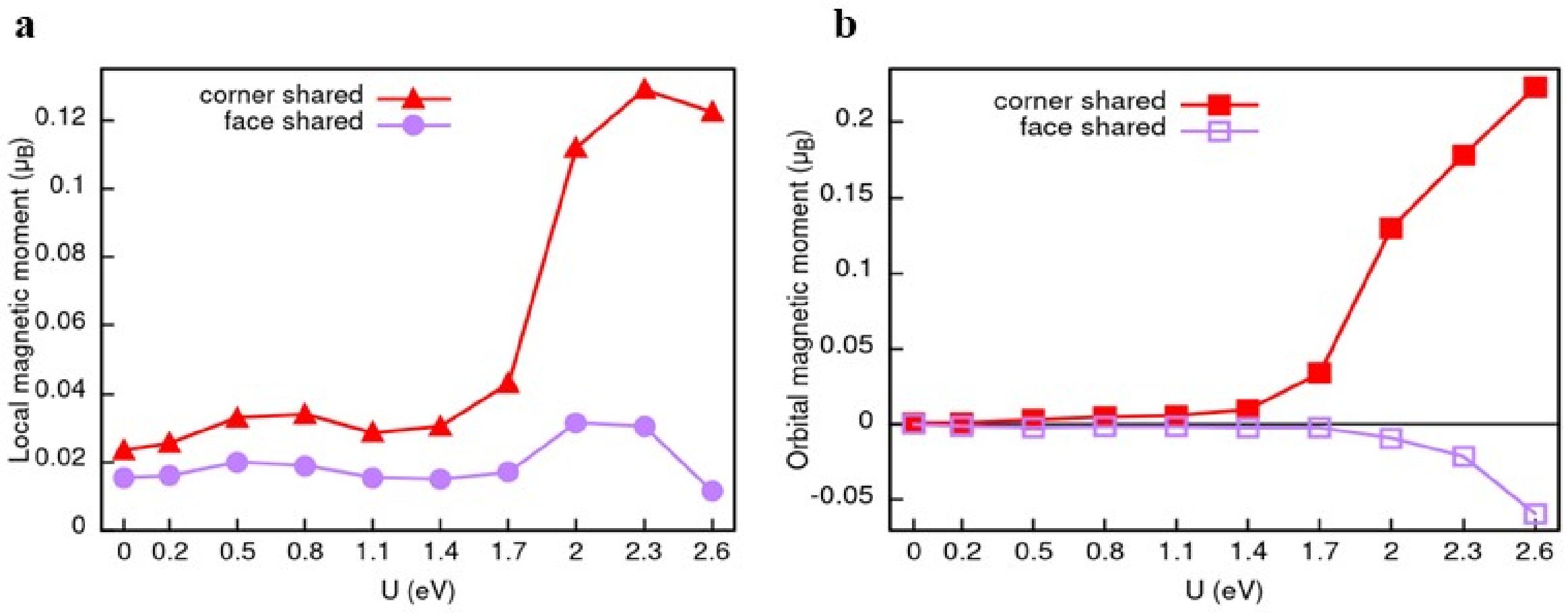

**Figure S8. a,** Local spin magnetic moment and **b**, orbital magnetic moment of SIO (111) as a function of the Coulomb repulsion. The results show a small spin magnetic moment for both corner-shared and face-shared octahedra. At low U, the orbital magnetic moment is small but the orbital magnetic moments show bigger values than the spin magnetic moment at high values of U.

The SIO (111) oriented structure was constructed starting from the hexagonal unit cell with one Ir atom per layer and lattice constants given from experimental measurements. The crystal structure studied hosts both $IrO_6$ corner-shared and face-shared octahedra with volumes 10.9 $Å^3$ and 10.6 $Å^3$, respectively. These two Ir sites strongly differ in the crystal, electronic, and magnetic properties. In order to consider different magnetic orders including antiferromagnetic phases, we have doubled the number of Ir atoms per layer and studied the orthorhombic supercell which has the same properties of the hexagonal one.

The ground state is the ferromagnetic phase for SIO with both corner-shared and face-shared octahedra and has its easy axis along the z-direction as shown in Fig. 4a of the main manuscript. From Fig. 4d of the main manuscript, it is evident from the peaks at the Fermi level that the ferromagnetic SIO is of the itinerant type and can be understood within a simple Stoner-type framework. In fact, our band structure calculations yield a finite density of states $D(E_F)$ at the Fermi level, which, when combined with the effective on-site Coulomb interaction U, satisfies the Stoner criterion $D(E_F)\cdot U>1$. This indicates that the system is unstable toward a spin-polarized state, thereby favoring itinerant Stoner ferromagnetism. In Fig. S8, we study the evolution of the moments as a function of the Coulomb repulsion U. We analyze spin and orbital magnetic moment respectively in Figure S8a and S8b. The corner-shared spin moment is always larger than the face-shared spin moment, this derives from the larger DOS at the Fermi level of the corner-shared octahedra which are the driving orbitals in the Stoner instability. Indeed, the Stoner instability can be orbital selective in the presence of multi-orbitals [8]. In the region which is physically relevant around U=0.8 eV, the spin moment is larger than the orbital moment. The orbital moments of the two distinct Ir atoms are always opposite, even if they are negligible for low values of U.

## S9-10. Band structure and Fermi surface for SIO (111)

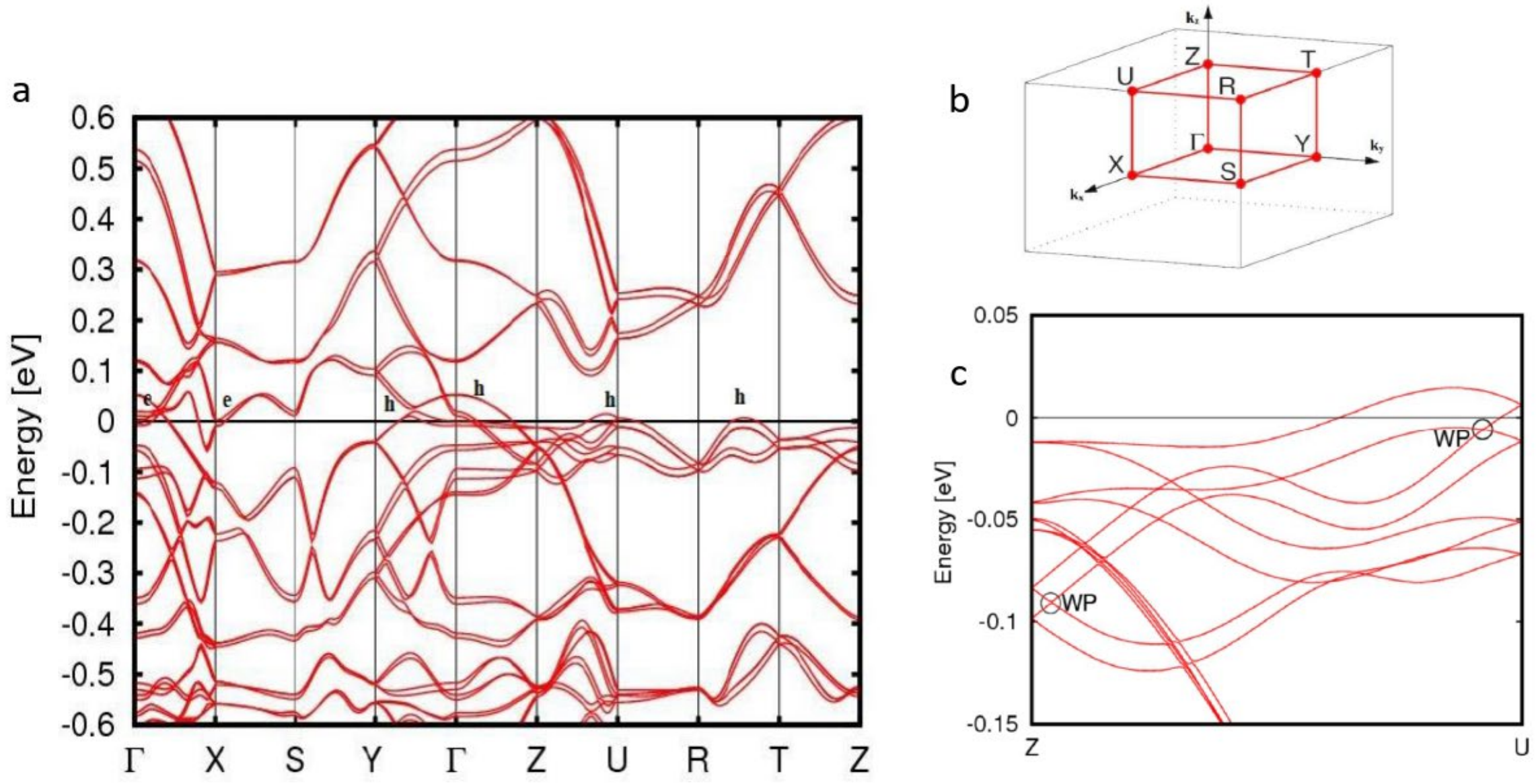

**Figure S9. a,** Band structure of SIO (111) with face-shared and corner-shared octahedra including SOC for U = 0.8 eV. **b,** High-symmetry points of the Brillouin zone. **c,** Magnetification of the band structure along the Z-U path. Close to the black circle, there are Weyl points (WP) arising from the crossing of semi-Dirac points.

We performed band structure calculations for SIO with both corner-shared and face-shared octahedra, shown in Fig. S9a along the k-path depicted in Fig. S9b. The system hosts non-symmorphic symmetries that produce semi-Dirac points at specific high-symmetry locations [9] without magnetization. In the presence of a small magnetization, the semi-Dirac points with spin-up and spin-down split in energies. The crossing of these semi-Dirac points creates Weyl points (WP) that enhance the AHE [10, 11]. WPs are created in positions of the k-space close to the black circles marked in Fig. 9c. In the figure, it is also visible from the band structure that there are more holes (h) than electrons (e) near the Fermi level, which indicates that the charge transport is dominated by holes. This can be also seen in the figures of the Fermi surfaces in Fig. S10.

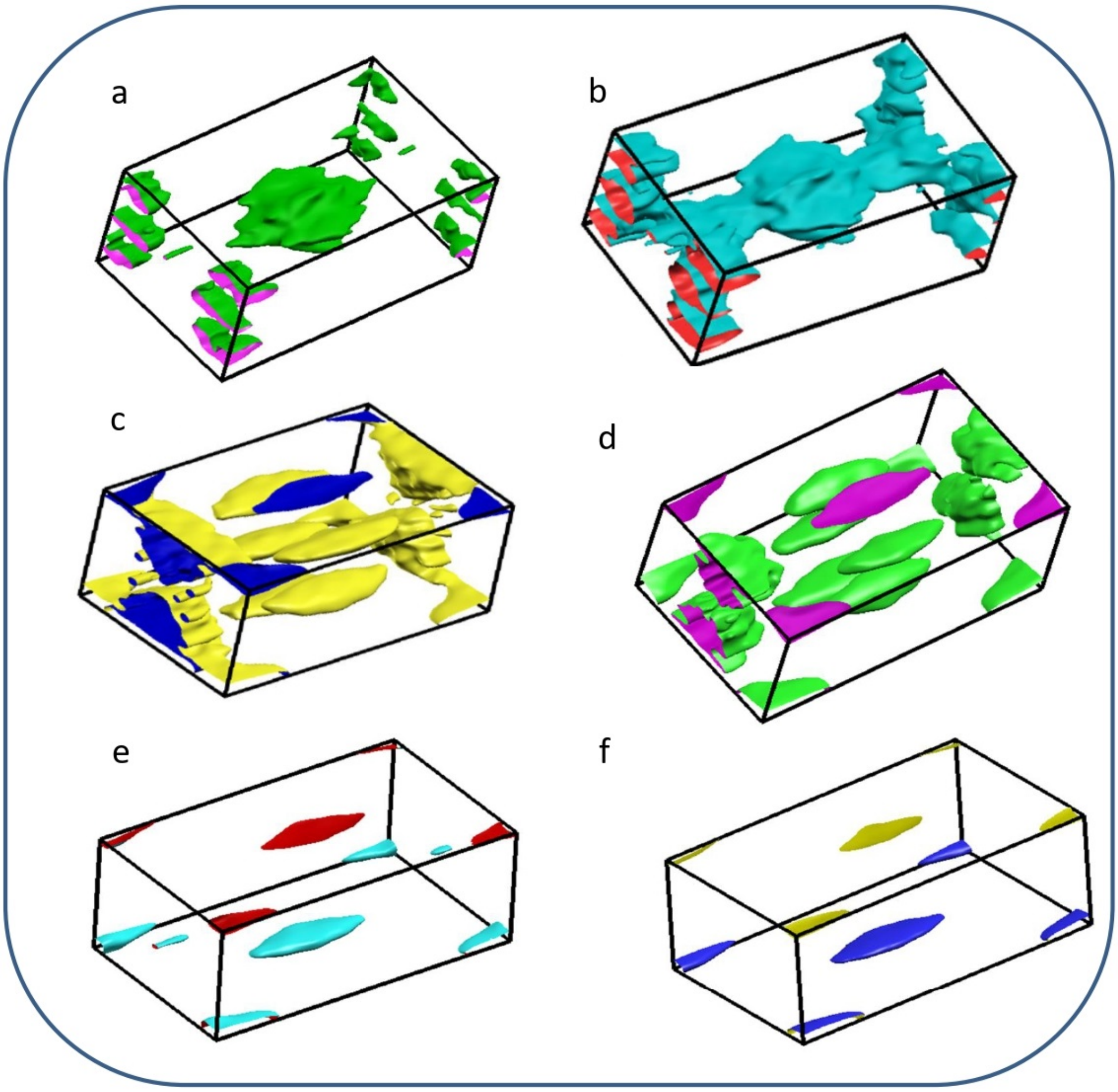

**Figure S10** Fermi surfaces of SIO (111) in GGA+SOC+U for U=0.8 eV. The system hosts 6 inequivalent Fermi surfaces which are labelled **a, b, c, d, e** and **f**.

Figure S10 shows the 6 independent Fermi surfaces of SIO (111). In the limit of zero magnetic moment, we would have Kramers degeneracy and three Fermi surfaces which are doubly degenerate. We can observe the pairs of Fermi surfaces that would be degenerate in the zero moment limit, these pairs are S10a and S10b, S10c and S10d and finally S10e and S10f. The Fermi surfaces have constraints close to the zone boundary in the presence of non-symmorphic symmetries [9]. There are more hole pockets than there are electron pockets in agreement with the band structure calculations. The self-consistent calculations were first done using a 10x10x10 k-points grid and after that with a 20x20x20 k-points grid for the non self-consistent calculation.

## References

[1] A. Biswas, K.-S. Kim and Y. H. Jeong, *J. Magn. Magn. Mater.* **2016**, 400, 36–40.

[2] L. Zhang, B. Pang, Y. B. Chen and Y. Chen, *Crit. Rev. Solid State Mater. Sci.* **2018**, 43, 367–391.

[3] Y. F. Nie, P. D. C. King, C. H. Kim, M. Uchida, H. I. Wei, B. D. Faeth, J. P. Ruf, J. P. C. Ruff, L. Xie, X. Pan, C. J. Fennie, D. G. Schlom and K. M. Shen, *Phys. Rev. Lett.* **2015**, 114, 016401.

[4] A. Tan, V. Labracherie, N. Kunchur, A. U. B. Wolter, J. Cornejo, J. Dufouleur, B. Büchner, A. Isaeva, R. Giraud, *Phys. Rev. Lett.* **2020**, 124, 197201.

[5] H. Unoki and T. Sakudo, *J. Phys. Soc. Jpn.* **1967**, 23, 546.

[6] R. Loetzsch, A. Lübcke, I. Uschmann, E. Förster, V. Große, M. Thuerk, T. Koettig, F. Schmidl, and P. Seidel, *Appl. Phys. Lett.* **2010**, 96, 071901.

[7] D. J. Groenendijk, N. Manca, G. Mattoni, L. Kootstra, S. Gariglio, Y. Huang, E. van Heumen and A. D. Caviglia, *Appl. Phys. Lett.* **2016**, 109, 041906.

[8] G. Cuono, C. Autieri and M. Wysokiński, *Phys. Rev. B* **2021**, 104, 024428.

[9] G. Cuono, F. Forte, M. Cuoco, R. Islam, J. Luo, C. Noce and C Autieri, *Phys. Rev. Mat.* **2019**, 3 (9), 095004.

[10] N. Mohanta, J. M. Ok, J. Zhang, H. Miao, E. Dagotto, H. N. Lee and S. Okamoto, *Phys. Rev. B* **2021**, 104, 235121.

[11] W. Meng, Y. Liu, X. Zhang, Z. He, W.-W. Yu, H. Zhang, L. Tian and G. Liu, *Phys. Rev. B* **2023**, 107, 115167.